\documentclass[a4paper,12pt]{article}

\usepackage{url}
\usepackage{amsmath,amsfonts,mathtools}
\usepackage[final]{graphicx}
\usepackage{color}
\usepackage{lmodern}
\usepackage{verbatim}
\usepackage{bbold}
\usepackage{booktabs}
\usepackage{bm}
\usepackage{amssymb}
\usepackage{ae}
\usepackage{aecompl}
\usepackage[rftl]{floatflt}
\usepackage[margin=2em, font={small},labelfont={sc}]{caption}
\usepackage{vwcol} 
\usepackage{multirow,bigdelim,dcolumn}
\allowdisplaybreaks  

\newcommand{\be}{\begin{equation}}
\newcommand{\ee}{\end{equation}}
\newcommand{\bdm}{\begin{displaymath}}
\newcommand{\edm}{\end{displaymath}}
\newcommand{\bea}{\begin{eqnarray}}
\newcommand{\eea}{\end{eqnarray}}
\newcommand{\bpm}{\begin{pmatrix}}
\newcommand{\epm}{\end{pmatrix}}
\newcommand\lsim{\mathrel{\rlap{\lower4pt\hbox{\hskip1pt$\sim$}}
    \raise1pt\hbox{$<$}}}
\newcommand\gsim{\mathrel{\rlap{\lower4pt\hbox{\hskip1pt$\sim$}}
    \raise1pt\hbox{$>$}}}
\DeclareMathOperator{\Tr}{Tr}

\newcommand{\TeV}{\, \mathrm{TeV}}
\newcommand{\GeV}{\, \mathrm{GeV}}

\newcommand{\MGUT}{M_{\rm GUT}}
\newcommand*{\tran}{^{\mkern-1.5mu\mathsf{T}}}  

\newcommand*\xbar[1]{%
   \hbox{%
     \vbox{%
       \hrule height 0.5pt 
       \kern0.5ex
       \hbox{%
         \kern-0.1em
         \ensuremath{#1}%
         \kern-0.1em
       }%
     }%
   }%
}

\newcommand{\SU}[1]{\ensuremath{{\text{SU}(#1)}}}

\newcommand{\U}[1]{\ensuremath{{\text{U}(#1)}}}

\newcommand{\irrepbase}[2][0]{\ensuremath{\text{\boldirrep{#2}}}}

\newcommand{\boldirrep}{\textbf}
\newlength{\irrepwidth}
\newlength{\irrepbarthickness}
\newlength{\irrepbarheight}
\newcommand{\irrepbarbase}[1]{%
    \settoheight{\irrepbarheight}{\ensuremath{\text{\boldirrep{#1}}}}%
    \settowidth{\irrepwidth}{\ensuremath{\text{\boldirrep{#1}}}}%
    \makebox[0pt][l]{\ensuremath{\text{\boldirrep{#1}}}}%
    \rule[1.2\irrepbarheight]{\irrepwidth}{\irrepbarthickness}%
}

\def\primes#1#2{\count0=#1 \loop \ifnum\count0>0 \advance\count0 by -1 #2\repeat}
\newcommand{\irrep}[2][0]{\ensuremath{\irrepbase{#2}^{\primes{#1}{\prime}}}}
\newcommand{\irrepbar}[2][0]{\ensuremath{\irrepbarbase{#2}^{\primes{#1}{\prime}}}}
\newcommand{\irrepsub}[3][0]{\ensuremath{\irrep[#1]{#2}_{#3}}}
\newcommand{\irrepbarsub}[3][0]{\ensuremath{\irrepbar[#1]{#2}_{#3}}}%
\newcommand{\SM}{$\,\SU{3}_c \times \protect\linebreak[0]\SU{2}_L \times \protect\linebreak[0]\U{1}_Y\,$}

\newcounter{comment}

{\refstepcounter{comment}%
\begin{quote}
\ttfamily\small$\blacksquare$ \textbf{\underline{Comment} $\sharp$\thecomment:}}%
{\end{quote}}

\def\TABSPACE{0.2em}

\begin{document}
\hfill
\begin{minipage}{20ex}\small
ZTF-EP-17-07\\
\end{minipage}
\begin{center}
{\LARGE \bf 
Renormalizable SU(5) Completions\\ of a Zee-type Neutrino Mass Model\\}
\vspace{0.7in}
{\bf  
Kre\v{s}imir~Kumeri\v{c}ki, Timon~Mede, Ivica~Picek}\\[4ex]
\begin{flushleft}
{\sl
Department of Physics, Faculty of Science, University of Zagreb,
 P.O.B. 331, HR-10002 Zagreb, Croatia\\}
\end{flushleft}

\vspace{0.5in}
\today \\[5ex]
\end{center}

\vspace{0.2in}

\begin{abstract}
We explore the potential of a selected model of radiative neutrino masses
to be implemented in a renormalizable \SU{5} unification framework.
The Zee-type model under consideration uncovers the \SU{5} representations in which the new fields are embedded and which may contain also other light states leading to the unification of gauge couplings. We perform an exhaustive search which reveals specific patterns of new states and demonstrate that such patterns are consistent with a general choice of relevant scalar potential. 
It turns out that all of the specific scenarios which lead to successful unification include the colored scalars testable at the LHC.

\end{abstract}
\vspace{0.2in}

\begin{flushleft}
\emph{Keywords}: Neutrino mass; Grand unification
\end{flushleft}

\clearpage

\section{Introduction}

Despite all the phenomenological success of the Standard Model (SM), certain
theoretical and experimental issues like neutrino masses, dark matter, charge
quantization, hierarchy problem etc. seem to indicate the need to go beyond its
well-established framework.
The ultraviolet completions motivated by neutrino mass models may address the
open questions and pave the road beyond the Standard Model (BSM). For example,
the neutrino masses in canonical type-I~\cite{Minkowski:1977sc,Yanagida:1979as,GellMann:1980vs,Glashow:1979nm,Mohapatra:1979ia}, type-II~\cite{Konetschny:1977bn,Magg:1980ut,Schechter:1980gr,Cheng:1980qt,Lazarides:1980nt,Mohapatra:1980yp}, and type-III~\cite{Foot:1988aq} tree-level seesaw models percolate down from a single scale that may be
linked to the unification point of SM gauge couplings, hinted first within the
\SU{5} Grand Unified Theory (GUT) of Georgi and
Glashow~\cite{Georgi:1974sy}.
After realizing that there is no single gauge coupling crossing in this
simplest GUT, it was noticed that augmenting  the SM  by the second Higgs
doublet and the corresponding supersymmetric (SUSY) partners enables
a successful minimal SUSY SM (MSSM) unification~\cite{Amaldi:1991cn}. 
A decisive role~\cite{Li:2003zh} played by incomplete (or {\em split})  
irreducible representations ({\em irreps}) \irrepsub{5}{H} in the MSSM unification success, motivated the corresponding non-SUSY attempts to cure the crossing problem~\cite{Krasnikov:1993sc,Willenbrock:2003ca} with
just six copies of the SM Higgs doublet field and nothing more. Still, 
the scale of such unifications would be too low.
 
Further studies of unification in the context of non-SUSY \SU{5} GUTs employed
incomplete \SU{5} irreps which contain new states introduced by
tree-level seesaw models. The studies
in~\cite{Bajc:2006ia,Bajc:2007zf,Dorsner:2006fx} employed 
adjoint \SU{5} representation \irrepsub{24}{F} which contains both the fermion
singlet and the $\TeV$-scale fermion triplet fields
providing a low scale hybrid of type-I and type-III seesaw models. Similarly, Refs.~\cite{Dorsner:2005ii,Dorsner:2005fq,Dorsner:2007fy} employed 
\irrepsub{15}{S} \SU{5} representation with the $\TeV$-scale complex scalar triplet, employed in the type-II seesaw mechanism. 

When considering possible GUT-embeddings of a radiative neutrino mass generating mechanism, we opt for genuine radiative Zee-type models, 
{\em genuine} in the sense that no additional symmetries are required to make them the dominant contribution to neutrino mass. At the same time, by avoiding fermion singlets we are choosing the \SU{5} embedding and discard the $SO(10)$ one. 
The first one-loop model proposed by Zee~\cite{Zee:1980ai} has introduced only
new \emph{scalar} fields, the charged singlet and the second complex doublet, which do not lead to competing tree-level seesaw mechanisms.  The embedding of original Zee model in renormalizable non-SUSY  \SU{5} setup has been studied in~\cite{Perez:2016qbo}. 
 
Our focus here will be on the variant of the Zee model presented in~\cite{Brdar:2013iea}, which  in the following we call the BPR model. It keeps the Zee's  charged scalar singlet, but a real scalar triplet replaces Zee's second Higgs doublet. Finally, BPR model introduces three copies of vector-like lepton doublet fields which, if embedded in split \irrepsub{5}{F}, may influence the gauge running as twelve Higgs doublets. 
Let us note that besides the genuine one-loop model~\cite{Brdar:2013iea} there exist three-loop  radiative neutrino models~\cite{Culjak:2015qja,Ahriche:2015wha}, where an automatic 
protection from the tree-level or lower loop-contributions has been achieved by introducing appropriate larger weak multiplets. 
However, the appearance of $\sim 10^6\GeV$ Landau pole (LP) for the $\SU{2}_L$ gauge coupling $g_2$ \cite{Sierra:2016qfa} eliminates these models from a unification framework. In contrast, as demonstrated in~\cite{Antipin:2017wiz}, the BPR one-loop model with the scalar triplet as the largest weak representation exhibits, in addition to the absence of LP, perturbativity and stability up to the Planck scale.
Therefore, we proceed here with the study of the gauge coupling unification in the context of the BPR loop model~\cite{Brdar:2013iea} for which the 
above requirements with respect to Yukawa and quartic couplings may remain valid when including extra color octet or color sextet scalar fields~\cite{Heikinheimo:2017nth}.
As it will turn out, adding these fields may be crucial to achieve the proper gauge unification. 

In Sec.~\ref{sec:BPR} we first present the set of BSM particles from the neutrino model~\cite{Brdar:2013iea}, dubbed BPR particles, and then present the gauge-unification conditions which the newly introduced states have to satisfy. In Section~\ref{sec:completions}, we will study the conditions under which the gauge couplings unify, and the particle spectra which make a proper unification possible.
Then in Section~\ref{sec:spectrum} we will show that the appropriate particle spectra are consistent with the scalar potential of our \SU{5} GUT scenarios.
We conclude in Sec.~\ref{sec:Conclusions}. The details of the algorithm for search are given in App.~A and the details of \SU{5} representations in App.~B.

\section{BPR model from GUT perspective}\label{sec:BPR}

\subsection{BPR-model states}\label{sec:BPR-model}

\begin{figure}[t]
\centering
\centerline{\includegraphics[scale=0.45]{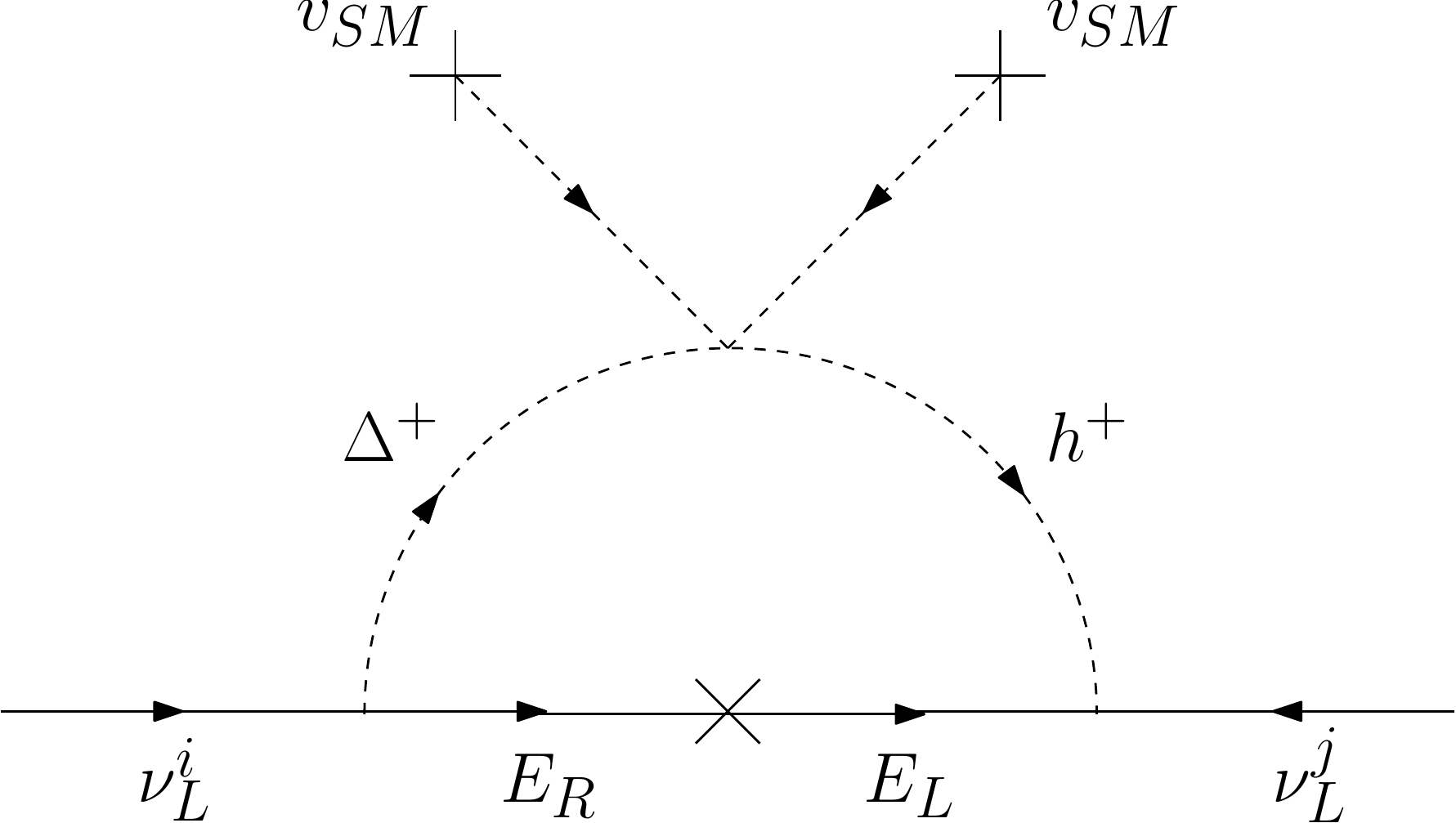}}
\caption{The one-loop BPR \protect\cite{Brdar:2013iea} neutrino mass mechanism.}
\label{fig:bpr}
\end{figure}

We adopt a simple and predictive TeV-scale radiative model~\cite{Brdar:2013iea} in which the loop contribution is genuine, i.e. self-protected like in the original Zee model. In its present variant the color singlet, weak triplet, hypercharge zero scalar field
$\Delta \sim (1,3,0)$,
\begin{equation}
\Delta=\frac{1}{\sqrt{2}}\sum_{j}\sigma_{j}\Delta^{j}=
    \left(  \begin{array}{ccc}
       \frac{1}{\sqrt{2}} \Delta^0 & \Delta^+\\
       \Delta^- & -\frac{1}{\sqrt{2}} \Delta^0
    \end{array} \right) \; ,
\end{equation}
is supplemented by a charged scalar singlet
\begin{equation}
    h^+ \sim (1,1,1)\ ,
\end{equation}
and by additional three generations of vector-like lepton doublets
\begin{equation}
    E_R \equiv (E_R^0, E_R^-)^T \sim (1,2,-1/2) \ , \ \ E_L \equiv (E_L^0,E_L^-)^T \sim (1,2,-1/2)\ ,
\end{equation}
which are needed to close the neutrino mass loop diagram displayed
in Fig.~\ref{fig:bpr}.
The corresponding vertices in the loop diagram are provided by Yukawa and quartic couplings in
\begin{equation}
-\mathcal{L} \supset  y_1 \overline{(L_L)^c} E_L h^+
+ y_2 \overline{L_L} \Delta E_R  + \lambda_7 H^\dag\Delta H^c h^+ + \mathrm{h.c.} \ .
\end{equation}
The vacuum expectation value $v_{\rm SM}$ of the SM Higgs doublet $H$ leads to the neutrino mass matrix
\begin{equation}
\mathcal{M}_{ij}=\sum_{k=1}^3\frac{[(y_1)_{ik} (y_2)_{jk} + (y_2)_{ik}(y_1)_{jk}]}
{8\pi^{2}} \ \lambda_7 \: v_{\rm SM}^2 \: M_{E_k} \; f(M_{E_k}, m_{\Delta^+}, m_{h^+}) \; ,
\label{effective}
\end{equation}
where $f(m_1, m_2, m_3)$ is a loop function specified in \cite{Brdar:2013iea}.
Assuming like in~\cite{Brdar:2013iea} the mass values $M_E \sim
m_{\Delta^+} \sim m_{h^+} \sim 200\!-\!500$ GeV, Eq.~(\ref{effective}) 
leads to $m_\nu \sim 0.1\,\mathrm{eV}$
for the couplings $y_{1,2}$ and $\lambda_7$ of the order of $10^{-4}$.
For definiteness, we will in the most of this work keep masses of these
new states fixed at $500 \GeV$.
In principle, even much larger masses would lead to a viable
neutrino mass model, with larger but still perturbative values of
$y_{1,2}$ and $\lambda_7$. Still, as we shall discuss later, such scenarios
would not bring much additional insight from the GUT perspective.

We display in Table~\ref{tab:SU(5)-particle-content} the  \SU{5} embedding
of the SM extended by states in the neutrino mass model at hand.  We note that additional potentially light BPR particles are described by the same SM group representations as those already populated by the SM states: new vector-like fermionic doublets  $E_{R,L}$ belong to the same representation as the Higgs $H^c$, 
and similarly for the charged scalar singlet $h^+$ and
the SM lepton singlet $e^{c}_R$, or the scalar adjoint triplet
$\Delta$  and the spin one triplet $W_{\mu}^i$. 
Understanding the quantum numbers of SM particles was one of the main 
motivations which led to the development of GUTs. 
The fact that BPR states populate already established SM
representations could be viewed as an additional motive to study them 
in the GUT setup.

\subsection{Matching BPR states with SU(5) irreps}
\label{sec:BPR-embedding}

One of the strongest arguments in favor of original Georgi-Glashow GUT scenario
is a neat embedding of all SM fermion representations, with apparently
arbitrary quantum numbers, into sum $\irrepbarsub{5}{F}\oplus \irrepsub{10}{F}$ of 
just two complete lowest \SU{5} representations. 
Since the gauge bosons have to belong to adjoint multiplet of the  \SU{5} group, 
essentially the only remaining unknown has been the structure of the scalar sector. 
In the present study, this 
generalizes to question of incorporating the well motivated BPR
set of particles into the same GUT context.

Following the pattern of SM states, and general principles of
economy and elegance, BPR states at hand may be
expected to belong to the lowest possible representations of the SU(5) group.
This would put scalar $\Delta$ in an adjoint \irrep{24}, scalar $h^+$ in
\irrep{10}, and vector-like leptons $E_{R,L}$ in appropriate
number of \irrep{5}+\irrepbar{5}, which is a choice displayed
in the right column of Table~\ref{tab:SU(5)-particle-content}.
These new \SU{5} irreps contain additional states displayed
in Table~\ref{tab:SU(5)-particle-content}, that are not needed for BPR neutrino mass mechanism. Some among these additional states
will prove crucial in obtaining the desired gauge coupling unification.

\begin{table}[t]
\caption{Particle content and the 
\SU{5} embedding options of the
BPR neutrino mass model~\cite{Brdar:2013iea}.
}
\label{tab:SU(5)-particle-content}
\centering
\vskip 0.2cm
\renewcommand{\arraystretch}{1.2}
\begin{tabular}{*{3}{c}}
\toprule
& SM + BPR 
& $\subset SU(5)$ \\
\addlinespace[\TABSPACE]\midrule
scalar & $H=(1,2,+\tfrac{1}{2})$ & $\irrep{5}=(1,2,+\tfrac{1}{2})\oplus(3,1,-\tfrac{1}{3})$ ; or \irrep{45}, \irrep{70} \\\addlinespace[\TABSPACE]
\cline{2-3} \addlinespace[\TABSPACE]
& $\Delta=(1,3,0)$ & 
$ \begin{aligned}[m] \irrep{24} &= (1,3,0)\oplus(8,1,0)\oplus (1,1,0) \\
    &\qquad{}\oplus(3,2,-\tfrac{5}{6})\oplus(\xbar{3},2,+\tfrac{5}{6})
\end{aligned} $ \\[3ex]
& $h^+=(1,1,+1)$ & $\irrep{10}=(1,1,+1)\oplus(\xbar{3},1,-\tfrac{2}{3})\oplus (3,2,+\tfrac{1}{6})$ \\
\addlinespace[\TABSPACE]\midrule
fermion & $3\times Q=(3,2,+\tfrac{1}{6})$ & 
\multirow{3}{*}{$3\times \irrep{10} = (3,2,+\tfrac{1}{6})\oplus (\xbar{3},1,-\tfrac{2}{3})\oplus (1,1,+1)$} \\
& $3 \times u^c=(\xbar{3},1,-\tfrac{2}{3})$ & \\
& $3 \times e^c=(1,1,+1)$ & \\
& $3\times L=(1,2,-\tfrac{1}{2})$ & 
\multirow{2}{*}{$3\times \irrepbar{5} = (1,2,-\tfrac{1}{2})\oplus (\xbar{3},1,+\tfrac{1}{3})$} \\
& $3 \times d^c=(\xbar{3},1,+\tfrac{1}{3})$ & \\\addlinespace[\TABSPACE]
\cline{2-3}
\addlinespace[\TABSPACE]
& $3 \times E_{R,L}=(1,2,-\tfrac{1}{2})$ & $3\times \irrepbar{5}= (1,2,-\tfrac{1}{2})\oplus (\xbar{3},1,+\tfrac{1}{3})$ ; or \irrepbar{45}, \irrepbar{70} \\
\addlinespace[\TABSPACE]\midrule
gauge & $G_{\mu}=(8,1,0)$ & 
\multirow{3}{*}{$ \begin{aligned}[m] \irrep{24} &= (1,3,0)\oplus(8,1,0)\oplus (1,1,0) \\
    &\qquad{}\oplus(3,2,-\tfrac{5}{6})\oplus(\xbar{3},2,+\tfrac{5}{6})
\end{aligned} $} \\
& $W_{\mu}=(1,3,0)$ & \\
& $B_{\mu}=(1,1,0)$ & \\
\addlinespace[\TABSPACE]
\bottomrule
\end{tabular}
\end{table}

To completely specify the structure of our model Lagrangian, we need to choose
the \SU{5} irreps that will contain the Standard Model Higgs $H$.
Here the most economical choice would be \irrep{5}, but, as we
shall see, this would not lead to a viable GUT model. 
Thus, we will consider also the options where the Higgs state 
belongs to \irrep{45} or \irrep{70} 
irreps\footnote{Using even higher \SU{5} irreps would expose us to the danger of low Landau poles.}, and we complete $\SU{5} \supset $ \SM branching
rules from Table~\ref{tab:SU(5)-particle-content} with:
\begin{align}
\irrep{45} = & \:(1,2,+\tfrac{1}{2}) \oplus (3,1,-\tfrac{1}{3}) \oplus (3,3,-\tfrac{1}{3}) 
\oplus (\xbar{3},1,+\tfrac{4}{3}) \oplus (\xbar{3},2,-\tfrac{7}{6}) \,\oplus \nonumber \\ 
& \oplus (\xbar{6},1,-\tfrac{1}{3}) \oplus (8,2,+\tfrac{1}{2}) \; , \label{eq:45}\\
\irrep{70} = & \:(1,2,+\tfrac{1}{2}) \oplus (1,4,+\tfrac{1}{2}) 
\oplus (3,1,-\tfrac{1}{3}) \oplus (3,3,-\tfrac{1}{3}) 
\oplus (\xbar{3},3,+\tfrac{4}{3}) \, \oplus \nonumber \\ 
& \oplus (6,2,-\tfrac{7}{6}) \oplus (8,2,+\tfrac{1}{2}) \oplus (15,1,-\tfrac{1}{3}) \; .
\label{eq:70}
\end{align}
As is known, Higgs in a mixture of \irrep{5} and
\irrep{45} can improve the GUT
fermion mass relations, like in the  
Georgi-Jarlskog mechanism~\cite{Georgi:1979df}.
However, in the present work we will not study the pattern of SM fermion masses.

On the other hand, a pattern of scalar masses and of new vector-like lepton masses \emph{is} important for our considerations, since these particles decisively affect the running of gauge couplings.
We need then to check for any of these possibilities whether unification can be achieved for large enough $M_{\rm GUT}$, whether scalar Lagrangian at renormalizable level allows for required masses of particles and, finally, whether such scenario is in compliance with phenomenological constraints 
and general theoretical requirement of perturbativity~\cite{Heikinheimo:2017nth,Kopp:2009xt}.

\subsection{Gauge coupling unification criteria}\label{sec:unif-criteria}


The unification of gauge couplings is controlled by the renormalization group equations (RGE) which govern the running of gauge couplings with the one-loop $\beta$ coefficients given by
\begin{align}
b_i & = -\tfrac{11}{3} \sum_G T(G_i) D(G_i) + \tfrac{4}{3} \sum_F T(F_i) D(F_i) \kappa + \tfrac{1}{3} \sum_S T(S_i) D(S_i) \eta \; .
\end{align}
Here, the Dynkin indices $T(R_i)$ are defined as $T(R_i)\delta_{ab}\equiv \Tr[\hat{T}_a(R_i)\hat{T}_b(R_i)]$ for generators $\hat{T}_{a}$ in  gauge, fermion and scalar representations $G_i$, $F_i$ and  $S_i$, respectively, and are conventionally normalized to $\tfrac{1}{2}$ for fundamental representations of \SU{N} groups (and thus to $\tfrac{3}{5} \hat{Y}^2$ for $U(1)_Y$). $D(R_i) \equiv \prod_{j\neq i} dim(R_j)$, $\kappa$ being $\tfrac{1}{2}$ ($1$) for Weyl (Dirac) fermions and $\eta$ being $\tfrac{1}{2}$ ($1$) for real (complex) scalars.
The SM $\beta$ coefficients (including the Higgs doublet) are $b_i^{(SM)} = (\tfrac{41}{10},-\tfrac{19}{6},-7)$, and RGE have analytical solution
\begin{align}
\alpha_{i}^{-1}(\MGUT) & = \alpha_i^{-1}(m_Z)-\frac{1}{2\pi} B_i  \ln{\tfrac{\MGUT}{m_Z}} \;, 
\quad i = 1, 2, 3\;,
\label{eq:definitions}
\end{align}
with coefficients
\begin{equation}
    B_i = b_i^{\rm(SM)} + \sum_{m_k<\MGUT} \Delta b_i^{(k)} {r_k} \;,
    \label{eq:Bi}
\end{equation}
and threshold weight factor of BSM state $k$, defined as
\begin{equation}
    r_{k}  =  \frac{ \ln \MGUT/m_k }{\ln \MGUT /m_{Z}} \;,
    \label{eq:rk}
\end{equation}
with value between $1$ (for $m_k = m_Z$) and $0$ (for $m_k = M_{\rm GUT}$), depending on the mass of the BSM particle $m_k$.
The sum in (\ref{eq:Bi}) goes over all BSM states, with 
$\Delta b_{i}^{(k)} = b_{i}^{(k)} - b_{i}^{(k-1)}$ being the increase
in the $\beta$ coefficients at the threshold of a given BSM state,
and $b_{i}^{(0)} = b_{i}^{(\rm SM)}$. 

As first, the unification condition $\alpha_1(M_{\rm GUT}) = \alpha_2(M_{\rm GUT}) = \alpha_3(M_{\rm GUT}) \equiv \alpha_{\rm GUT}$ can be expressed in the form of the so-called B-test~\cite{Giveon:1991zm,Dorsner:2017ufx}:

\begin{align}
\frac{B_{23}}{B_{12}} & \equiv \frac{B_2-B_3}{B_1-B_2}=\frac{\alpha_2^{-1}(m_Z)-\alpha_3^{-1}(m_Z)}{\alpha_1^{-1}(m_Z)-\alpha_2^{-1}(m_Z)}=\tfrac{5}{8}\,\frac{\sin^2\theta_w(m_Z)-\tfrac{\alpha_{EM}(m_Z)}{\alpha_3(m_Z)}}{\tfrac{3}{8}-\sin^2\theta_w(m_Z)}=0.718 \;, 
\label{eq:Btest}
\end{align}
where we used the average numerical values for the constants at $m_Z$ scale,
as given in~\cite{Patrignani:2016xqp}.
The comparison to the corresponding SM value $0.528$ 
indicates that the couplings do not unify in the SM.

Second, the associated GUT scale 

\begin{align}
\MGUT & = m_Z \, \exp\left (\frac{2\pi (\alpha_1^{-1}(m_Z)-\alpha_2^{-1}(m_Z))}{B_1-B_2}\right ) = m_Z  \, \exp\left (\frac{184.87}{B_{12}}\right ) 
\label{eq:defMGUT}
\end{align}
yields for the SM the value $\MGUT = 10^{13}\GeV$. Therefore, additional BSM states should
improve unification and increase its scale up to at least $5\times 10^{15}\GeV$ 
which is in agreement with proton lifetime bounds \cite{Miura:2016krn}.
Such additional BSM states
must therefore provide a negative net contribution to $B_{12}$ and positive to
$B_{23}$.

\section{Possible gauge-unification realizations}\label{sec:completions} 

\subsection{Effect of BPR states on gauge unification}

Before embedding in the \SU{5} GUT framework, we first investigate how the new states, needed for neutrino mass mechanism, influence the RGE running and to what extent they alone could satisfy the unification criteria from Sect.~\ref{sec:unif-criteria}.

\begin{table}[th!]
\caption{Contributions of BPR states to RGE running, where
threshold weights $r_{k}$ are defined in Eq.~\eqref{eq:rk}. 
Note that two Weyl fermion states $E_{L,R}$ come in $N_{\rm vec}=3$ copies}
\label{tab:BPR-beta-coefficents}
\centering
\renewcommand{\arraystretch}{1.2}
\begin{tabular}{c c| c c}
 \multicolumn{2}{c|}{$k$} &  $\Delta B_{23}$  & $\Delta B_{12}$   \\ \hline
$h^+$ & $(1,1,1)$ & $0$& $\hphantom{+}\frac{1}{5} r_k$ \\
$\Delta$ & $(1,3,0)$ & $\hphantom{+}\frac{1}{3} r_k$& $- \frac{1}{3} r_k$ \\
$E_{L,R}$ & $(1,2,-1/2)$ & $\hphantom{+}\frac{1}{3} r_k$& $- \frac{2}{15} r_k$ \\
\end{tabular}
\end{table}

In Table~\ref{tab:BPR-beta-coefficents} we list extra BPR
states together with their contribution to pertinent combinations
of $\beta$-function coefficients. As already stressed, the states with positive
$B_{23}$ and negative $B_{12}$ are promising for unification. It can be readily
seen that only $h^+$ is not of this kind. 

If we consider the default configuration with all BPR states close
to the electroweak scale (i.e. weight factors from Eq.~\eqref{eq:rk} being 
$r_k \sim 1$), one immediately observes that the B-test combination
increases to $B_{23}/B_{12} = 0.974$, 
from the SM value $0.528$, considerably overshooting 
the required value of $0.718$ from Eq.~\eqref{eq:Btest}.
This is mostly due to the strong effect of three copies of BPR vector-like
lepton doublets. Since they actually double 
the RGE effect of previously mentioned six Higgs doublets,
one can similarly achieve correct unification if they are set at the
intermediate scale with the factor $r_k \sim 0.5$.
However, again like in the six Higgs doublet case, the unification
scale would be too low.

Indeed, one observes that there is no way to obtain high enough
unification scale by using only BPR states. Namely, even if negative effects
of scalar $h^+$ state are avoided by putting it at some very high
scale, the total contribution of $\Delta$ and $E_{L,R}$ states
to $\Delta B_{12}$ is at most only $-17/15$  (for $r_k = 1$)
resulting in $\MGUT=1.1\times 10^{15}\GeV$
as the maximal possible GUT scale, even if one would completely 
disregard the condition of the gauge coupling crossing.

In conclusion, BPR states alone cannot lead to a successful unification that
would at the same time respect proton decay bounds. To achieve that, some other
states below the GUT scale should 
be invoked. Such states are naturally provided by embedding BPR model in a \SU{5} unification framework, as we shall show in detail.
%

\subsection{Higgs doublet in \irrepsub{5}{H}}\label{sec:Higgs-in-5_H}

As explained in Sect.~\ref{sec:BPR-embedding}, we will first try
the simplest possible \SU{5} embedding of the BPR mechanism, where
SM Higgs doublet becomes a member of the \irrepsub{5}{H}.
The scalar sector of this model contains \irrepsub{5}{H}, 
\irrepsub{10}{S} and \irrepsub{24}{S} multiplets
(we use subscript $H$ on those scalar irreps of \SU{5} which contain
SM Higgs field), and there are $N_{\rm vec}$ 
generations of vector-like
matter in $\irrepsub{5}{F}\oplus\irrepbarsub{5}{F}$,
on top of the Standard Model quarks and leptons
in $n_{g}=3$ generations of \irrepsub{10}{F} and $\irrepbarsub{5}{F}$,
and gauge bosons in the adjoint \irrepsub{24}{g} representation, as displayed
in Table~\ref{tab:SU(5)-particle-content}. 
Their contributions to the RGE running are listed in
Table~\ref{tab:5_H-model-beta-coefficents}.

\begin{table}[th!]
\caption{BSM contributions to RGE running in the simplest \SU{5} embedding of the BPR mechanism. 
$H$ stands for SM Higgs doublet whose contribution has already been accounted
for by $b_i^{(SM)}$. The massless scalar leptoquarks $X$ and $Y$ get
absorbed into longitudinal components of massive gauge bosons, as dictated by the Nambu-Goldstone mechanism --- the
$\beta$ coefficients of these scalars thus enter at the same scale as heavy
vectors (\emph{i. e.} $r_k \approx 0$).}
\label{tab:5_H-model-beta-coefficents}
\centering
\renewcommand{\arraystretch}{1.2}
\begin{tabular}{c c c| c c}
 \multicolumn{3}{c|}{$k$}  &  $\Delta B_{23}$  & $\Delta B_{12}$ \\
\hline
$H$ & $(1,2,1/2)$ & \multirow{ 2 }{*}{ \irrepsub{5}{H} }& $\hphantom{+}\frac{1}{6} r_k$& $- \frac{1}{15} r_k$ \\
$S_1$ & $(3,1,-1/3)$ & & $- \frac{1}{6} r_k$& $\hphantom{+}\frac{1}{15} r_k$ \\
\hline
$h^+$ & $(1,1,1)$ & \multirow{ 3 }{*}{ \irrepsub{10}{S} }& $0$& $\hphantom{+}\frac{1}{5} r_k$ \\
 & $(\bar{3},1,-2/3)$ & & $- \frac{1}{6} r_k$& $\hphantom{+}\frac{4}{15} r_k$ \\
 & $(3,2,1/6)$ & & $\hphantom{+}\frac{1}{6} r_k$& $- \frac{7}{15} r_k$ \\
\hline
 & $(1,1,0)$ & \multirow{ 5 }{*}{ \irrepsub{24}{S} }& $0$& $0$ \\
$\Delta$ & $(1,3,0)$ & & $\hphantom{+}\frac{1}{3} r_k$& $- \frac{1}{3} r_k$ \\
 & $(8,1,0)$ & & $- \frac{1}{2} r_k$& $0$ \\
$X, Y$ & $(3,2,-5/6)$ & & $\hphantom{+}\frac{1}{12} r_k$& $\hphantom{+}\frac{1}{6} r_k$ \\
$\bar{X}, \bar{Y}$ & $(3,2,5/6)$ & & $\hphantom{+}\frac{1}{12} r_k$& $\hphantom{+}\frac{1}{6} r_k$ \\
\hline
$E_{L,R} $ & $(1,2,-1/2)$ & \multirow{ 2 }{*}{ \irrepbarsub{5}{F} }& $\hphantom{+}\frac{1}{3} r_k$& $- \frac{2}{15} r_k$ \\
$$ & $(\bar{3},1,1/3)$ & & $- \frac{1}{3} r_k$& $\hphantom{+}\frac{2}{15} r_k$ \\
\end{tabular}
\end{table}

It is known that the simplest Georgi-Glashow \SU{5} GUT suffers from
the doublet-triplet splitting problem, where leptoquark
$S_1 = (3, 1, -1/3)$, which completes \irrepsub{5}{H} together
with SM Higgs, has to be much heavier than the Higgs so that it
doesn't induce the fast proton decay. 
There is nothing preventing the other scalar \SU{5} multiplets to be split,
and we will check in Sect.~\ref{sec:spectrum} that our
splitting patterns are consistent with the structure of the general
scalar potential.
Still, whatever the actual mechanism responsible for the multiplet
splitting, there is no reason to assume that this mechanism
is somehow aligned with the neutrino mass mechanism. 
Thus, we will be quite general in allowing the splitting of masses
within \SU{5} multiplets.

With this freedom, and having at our disposal variety of states
from Table~\ref{tab:5_H-model-beta-coefficents} with different RGE behaviour,
the unification prospects look promising. Indeed, we have found several scenarios
where coupling constants correctly unify (cross at the single point).
However, we also find that, whatever the masses of BSM states between
$M_Z$ and $\MGUT$, the highest possible unification scale in this
model is $\MGUT < 10^{15}\,\GeV$, in violation of experimental bounds on 
proton decay widths. Thus, this simplest embedding of the proposed
neutrino mass model with the Higgs doublet restricted to \irrepsub{5}{H} irrep is ruled out.
%

\subsection{Higgs doublet in \irrepsub{45}{H}}\label{sec:Higgs-in-45_H}

Next we consider the scenario where the SM Higgs doublet is embedded into
\irrepsub{45}{H} instead of \irrepsub{5}{H}, or in some mixture
of both. The larger particle content can help raising the
unification scale and, as a bonus, this setup can serve to correct the wrong
mass relations between charged leptons and down-type quarks at the
renormalizable level which are usually obtained in the simplest \SU{5}
models. The $\beta$ coefficients of the extra states from scalar \irrepsub{45}{H} can be found in Table~\ref{tab:45_H-model-beta-coefficents}, which should
be added to states in Table~\ref{tab:5_H-model-beta-coefficents} to
obtain a complete embedding of the SM Higgs and the BPR states into \SU{5} multiplets.

\begin{table}[th!]
   \caption{Contributions of \irrepsub{45}{H} to running. For a complete model
  the multiplets from Table~\ref{tab:5_H-model-beta-coefficents} are to be
   added. The states $S_{1}$, $S_3$ and $\tilde{S}_1$ are leptoquarks that,
   if light, would induce too fast proton decay.
   }
\label{tab:45_H-model-beta-coefficents}
\centering
\renewcommand{\arraystretch}{1.2}
\begin{tabular}{c c c| c c}
 \multicolumn{3}{c|}{$k$}  &  $\Delta B_{23}$  & $\Delta B_{12}$ \\
\hline
$\Sigma_a$ & $(1,2,1/2)$ & \multirow{ 7 }{*}{ \irrepsub{45}{H} }& $\hphantom{+}\frac{1}{6} r_k$& $- \frac{1}{15} r_k$ \\
$S_1\equiv\Sigma_b$ & $(3,1,-1/3)$ & & $- \frac{1}{6} r_k$& $\hphantom{+}\frac{1}{15} r_k$ \\
$S_3\equiv\Sigma_c$ & $(3,3,-1/3)$ & & $\hphantom{+}\frac{3}{2} r_k$& $- \frac{9}{5} r_k$ \\
$\tilde{S}_1\equiv\Sigma_d$ & $(\bar{3},1,4/3)$ & & $- \frac{1}{6} r_k$& $\hphantom{+}\frac{16}{15} r_k$ \\
$\Sigma_e$ & $(\bar{3},2,-7/6)$ & & $\hphantom{+}\frac{1}{6} r_k$& $\hphantom{+}\frac{17}{15} r_k$ \\
$\Sigma_f$ & $(\bar{6},1,-1/3)$ & & $- \frac{5}{6} r_k$& $\hphantom{+}\frac{2}{15} r_k$ \\
$\Sigma_g$ & $(8,2,1/2)$ & & $- \frac{2}{3} r_k$& $- \frac{8}{15} r_k$ \\
\end{tabular}
\end{table}

In this more realistic model one finds many ways
in which one can achieve a correct unification, so we need to
specify some criteria that will lead to a set of models
covering all interesting scenarios; let us list
those implemented in our study.

\begin{itemize}
\item First, note that if all states of a given \SU{5} irrep appear at the
same mass scale, their effect on RGE cancels (contributions
to either $B_{23}$ or $B_{12}$ from all states of a given \SU{5} irrep in, 
\emph{e. g.}, Table~\ref{tab:5_H-model-beta-coefficents},
taking the same $r_k$, add up to zero). Thus, we will fix the
BPR states close to the electroweak scale (for definiteness,
we put them at $500\,\GeV$),
and by doing so we don't loose much generality, from the
standpoint of RGE, because the effect on
unfication of making e. g. BPR vector-like leptons $E_{L,R}$
heavier is the same as making the rest of the multiplet
(in this case $(3,1,-2/3)$ states) lighter. (Some generality
may be lost if some of these other states cannot be made lighter
for other reasons.)

\item Next, since, as discussed before, we allow general splitting
of \SU{5} multiplets, with any experimentally allowed mass for
the rest of BSM states (see Sect.~\ref{sec:spectrum}), we have
enough freedom to achieve the \emph{exact} gauge coupling unification
\emph{i.e.} a fulfilment of the B-test, see Eq.~\eqref{eq:Btest}.
Then, we require the GUT scale $\MGUT$ larger than $2\times 10^{15} \GeV$. 
The lowest experimental bound, coming from proton decay searches, is actually
about $5\times 10^{15}$; however, it turns out that for most of the scenarios
presented here, a simplified analysis (ignoring the Yukawa contributions to
2-loop RGE) shows that improving RGE to 2-loop accuracy increases $\MGUT$ beyond
$5\times 10^{15}\,\GeV$.

\item We will also exclude scenarios with very heavy new BSM particles,
    with masses between $\sim 10^{11} \GeV$
and $M_{\rm GUT}$.  Otherwise, one can always take any successful model, add
some particles slightly below $\MGUT$ that will have only small influence on
running, and thus obtain many more models which will be qualitatively the same
as the ones presented in this paper, only more complicated.
This requirement at the same time excludes from consideration
leptoquarks $S_1 = (3,1,-1/3)$, $\tilde{S}_{1} = (3,1,4/3)$ and $S_{3}=(3,3,-1/3)$  
which, if lighter than $\sim 10^{11} \GeV$ would naturally lead to proton decay
in violation of experimental limits.

\item For all BSM particles we take $500\,\GeV$ as a lower bound on their masses. Direct LHC searches
sometimes put higher bounds on such states, but these bounds are often obtained only within
specific benchmark scenarios. 
E. g. recent CMS search \cite{Sirunyan:2016iap} puts lower bound of $3\,\TeV$ on color
octet state, like $(8,1,0)$ from \irrepsub{24}{S}, 
but only within benchmark model of Refs. \cite{Han:2010rf,Chivukula:2014pma}, 
where couplings to loop fermions in production and decay are taken to be of order one (see also Ref. \cite{Hayreter:2017wra}).

\item We include only particles from single copies of scalar \SU{5}
irreps \irrepsub{5}{H}, \irrepsub{10}{S}, \irrepsub{24}{S} and
\irrepsub{45}{H} (or, in the next Sect. \irrepsub{70}{H}) 
and $N_{\rm vec} = n_{\rm g} = 3$ copies of vector-like \irrepbarsub{5}{F},
which are all already needed for embedding the BPR neutrino mass mechanism.
\end{itemize}

Under these conditions, we performed the exhaustive search of the parameter
space, using the algorithm specified in Appendix~\ref{app:algo}, and
resulting in successful scenarios listed in Table \ref{tab:H45}.
As explained in the Appendix~\ref{app:algo}, when a given set of new BSM states
offers a continuum of possible GUT scenarios (with different spectra), 
we represent this continuum by a specific choice of spectrum with
minimal average mass of particles\footnote{More precisely, maximal
average threshold weight factor $r_k$ defined in Eq.~\eqref{eq:rk}.}. Such
a choice is motivated, besides the need for definiteness, 
by the desire to focus on models which have maximal
discovery potential at LHC and future colliders.

\begin{table}
\renewcommand{\arraystretch}{1.2}
\caption{\label{tab:H45} Seven unification scenarios with SM Higgs in
${\mathbf{5}}$ and/or ${\mathbf{45}}$ of SU(5) and BPR states fixed at $\sim 0.5\TeV$.
  }
\begin{center}
\begin{tabular}{cc|ccccccc}
    \multicolumn{2}{c|}{irreps} & \multicolumn{7}{c}{$m_{k}$ [TeV]} \\
    SM & SU(5) &  A1  &  A2  & A3  & A4  &  A5  & A6  & A7 \\ \hline
    $(3,1, 1/3)$ & \irrepbarsub{5}{F} &  5000  &   & $2.3\times 10^6$  &
    450  &    & $2\times 10^5$  &  \\ \hline
    $(\bar{3},1,-2/3)$ & \multirow{2}{*}{\irrepsub{10}{S}} &  &  &  &   &  
    &      & 2.4 \\
    $(3,2, 1/6)$ &  & 0.5 &  & 0.5 & 0.5  &  &  0.5 & 0.5  \\ \hline
    $(8,1,0)$ & \irrepsub{24}{S} &  & 0.5 & 0.5 & & 0.5 & 0.5 &   \\ \hline
    $(1,2, 1/2)$ & \multirow{3}{*}{\irrepsub{45}{H}} &  &  &  & 0.5 &
     260 & 0.5  &          \\
    $(\bar{6},1, -1/3)$ & &  & 90  &  &  & 0.5 &  & 0.5\\ 
    $(8,2, 1/2)$  & & 0.5 & 0.5 & 0.5 & 0.5 & 0.5 & 0.5 & 0.5   \\ \hline
    \multicolumn{2}{c|}{$\MGUT^{\rm max}/(10^{15}\,\GeV)$} & 2.8 & 2.5 & 6.2 &
     2.8 & 2.8 & 6.2 & 6.5 
\end{tabular}
\end{center}
\end{table}

Note that a light $(8,2,1/2)$ is the only other allowed representation in $45_H$ with negative $B_{12}$ contribution needed to increase $M_{\rm GUT}$ and thus suppressing the proton decay. Of course by itself it doesn't help the unification due to negative $B_{23}$ (acting alone it can decrease ${B_{23}}/{B_{12}}$ to $0.470$),  but its strong effect on unification scale is important for
all models displayed in Table~\ref{tab:H45}.

%

\subsection{Higgs doublet in \irrepsub{70}{H}}\label{sec:Higgs-in-70_H}
If we opt for SM Higgs belonging to \irrepsub{70}{H} instead
of \irrepsub{45}{H} (in addition to \irrepsub{5}{H}),
the search proceeds under the conditions explicated in the previous
subsection,
and the $\beta$ coefficients of the extra states 
can be found in Table~\ref{tab:70_H-model-beta-coefficents}.

\begin{table}[th!]
  \caption{Contributions of \irrepsub{70}{H} to RGE running. For a complete model, the multiplets from Table~\ref{tab:5_H-model-beta-coefficents} are to be added. The states $S_{1}$ and $S_3$ are leptoquarks that,
   if light, would naturally induce too fast proton decay.
   }
\label{tab:70_H-model-beta-coefficents}
\centering
\renewcommand{\arraystretch}{1.2}
\begin{tabular}{c c c| c c}
 \multicolumn{3}{c|}{$k$}  &  $\Delta B_{23}$  & $\Delta B_{12}$ \\
\hline
$\Omega_a$ & $(1,2,1/2)$ & \multirow{ 8 }{*}{ \irrepsub{70}{H} }& $\hphantom{+}\frac{1}{6} r_k$& $- \frac{1}{15} r_k$ \\
$S_1 \equiv \Omega_b$ & $(3,1,-1/3)$ & & $- \frac{1}{6} r_k$& $\hphantom{+}\frac{1}{15} r_k$ \\
$S_3 \equiv \Omega_c$ & $(3,3,-1/3)$ & & $\hphantom{+}\frac{3}{2} r_k$& $- \frac{9}{5} r_k$ \\
$\Omega_d$ & $(\bar{3},3,4/3)$ & & $\hphantom{+}\frac{3}{2} r_k$& $\hphantom{+}\frac{6}{5} r_k$ \\
$\Omega_e$ & $(6,2,-7/6)$ & & $- \frac{2}{3} r_k$& $\hphantom{+}\frac{34}{15} r_k$ \\
$\Omega_f$ & $(15,1,-1/3)$ & & $- \frac{10}{3} r_k$& $\hphantom{+}\frac{1}{3} r_k$ \\
$\Omega_g$ & $(8,2,1/2)$ & & $- \frac{2}{3} r_k$& $- \frac{8}{15} r_k$ \\
$\Omega_h$ & $(1,4,1/2)$ & & $\hphantom{+}\frac{5}{3} r_k$& $- \frac{22}{15} r_k$ \\
\end{tabular}
\end{table}

In this setup, we find three
unification scenarios displayed in Table \ref{tab:H70} to which scenarios
A1, A3, A4 and A6 from Table \protect\ref{tab:H45} should be added,
since they employ only states from \irrepsub{45}{H} that
are also present in \irrepsub{70}{H}. 
From this search we have explicitly excluded representation $(15, 1, -1/3)$
which, if light, leads to Landau poles below $\MGUT$.
To avoid this, it should be heavier than at least $10^7\,\GeV$
\cite{Heikinheimo:2017nth} so that its effect on the RG running would be diminished.
Including also this representation leads to 16 additional scenarios beyond
those in Tables \ref{tab:H45} and \ref{tab:H70}, which we have chosen not to list.

\begin{table}
\renewcommand{\arraystretch}{1.3}
\caption{\label{tab:H70} Three unification scenarios with SM Higgs in
\irrepsub{5}{H} and \irrepsub{70}{H} of SU(5), and BPR states fixed at $\sim 0.5\TeV$.
These are at the same time \emph{all} viable scenarios 
under assumption of unsplit vector-like fermion \irrepsub{5}{F} .}
\begin{center}
\begin{tabular}{cc|ccc}
    \multicolumn{2}{c|}{irreps} & \multicolumn{3}{c}{$m_{k}$ [TeV]} \\
    SM & SU(5) &  B1  &  B2  & B3  \\ \hline
    $(3,1, 1/3)$ & \irrepbarsub{5}{F} &  0.5  & 0.5  & 0.5  \\ \hline
    $(3,2, 1/6)$ & \irrepsub{10}{S} & &   & 0.5 \\ \hline
    $(8,1,0)$ & \irrepsub{24}{S} &  & 0.5 & 0.5   \\ \hline
    $(1,4, 1/2)$ & \multirow{2}{*}{\irrepsub{70}{H}} & $1.8\times 10^6$ &
      $1.3\times 10^4$& $6.3\times 10^6$  \\
    $(8,2, 1/2)$ & & 0.5 & 0.5 & 0.5   \\ \hline
    \multicolumn{2}{c|}{$\MGUT^{\rm max}/(10^{15}\,\GeV)$} & 2.6 & 10.6 & 32.0
\end{tabular}
\end{center}
\end{table}

Interestingly, all viable scenarios in this setup, as displayed
in Table~\ref{tab:H70}, involve the color triplet $(3,1,1/3)$ at
the same scale ($500\,\GeV$) as BPR
vector-like leptons, making the irrep \irrepsub{5}{F} complete and nullifying its influence on the RGE running. Thus, these states
can all be at any other scale as well, without changing the gauge
unification property of the model.
We note that there are no 
viable scenarios with such unsplit fermion \irrepsub{5}{F} and
with Higgs in \irrepsub{45}{H}, \emph{i.e.} in the framework
of Sect.~\ref{sec:Higgs-in-45_H}.

\section{Scalar potential and spectrum}\label{sec:spectrum}

In previous Sec.~~\ref{sec:Higgs-in-45_H} and Sec.~\ref{sec:Higgs-in-70_H} we singled out viable scenarios in two variants of non-supersymmetric \SU{5}
unification. Now we are presenting for them the relevant expressions for the
scalar potentials and the resulting mass spectra, demonstrating their
consistency with the appropriate scalar sector extensions.
The scalar sector for the scenarios from Sec.~\ref{sec:Higgs-in-45_H} 
contains: 
\begin{itemize}
    \item \irrepsub{24}{S}: an adjoint $24$-dimensional real traceless 
    representation $\chi^{i}_{j}$; 
    \item  \irrepsub{5}{H}: a fundamental $5$-dimensional complex representation $H^{i}$; 
    \item \irrepsub{10}{S}: an antisymmetric  
    $10$-dimensional complex representation $\phi^{ij}$;
    \item \irrepsub{45}{H}: a $45$-dimensional complex $2$-index antisymmetric  
    traceless 
    representation $\Sigma^{ij}_{k}$,
\end{itemize} 
which get decomposed under the SM gauge group as displayed in 
Table~\ref{tab:SU(5)-particle-content} and Eq. (\ref{eq:45}).
For the scenario from Sec. \ref{sec:Higgs-in-70_H}, $\Sigma^{ij}_{k}$ is replaced with
\begin{itemize}
  \item \irrepsub{70}{H}: a $70$-dimensional complex $2$-index symmetric 
traceless 
representation $\Omega^{ij}_{k}$ ,
\end{itemize}
 with the SM decomposition displayed in Eq.~\eqref{eq:70}.
The details of individual representations can be found in the Appendix~\ref{app:irreps}.
The following fields from this set can develop potentially non-vanishing VEVs:
\begin{itemize}
\item the SM singlet field from $\chi^{i}_{j}$ whose GUT scale VEV $\langle (1,1,0)_{\chi} \rangle \equiv v_{GUT}$ results in breaking $SU(5) \to SU(3)_c \times SU(2)_L \times U(1)_Y$;
\item the neutral components of the weak doublets from $H^{i}$ and $\Sigma^{ij}_{k}$ whose $SU(2)_L \times U(1)_Y \to U(1)_{Q}$ breaking VEVs $\langle (1,2,+\tfrac{1}{2})_{H} \rangle \equiv v_{5}$ and $\langle (1,2,+\tfrac{1}{2})_{\Sigma} \rangle \equiv v_{45}$ are subject to the condition $v_{5}^2+v_{45}^2 = v_{SM}^2$;
\item the neutral components of $(1,2,+\tfrac{1}{2})_{\Omega}$ and $(1,4,+\tfrac{1}{2})_{\Omega}$ from $\Omega^{ij}_{k}$ can also develop VEVs of the order of electroweak scale;
\item the neutral component of the weak triplet from $\chi^{i}_{j}$ can develop a tiny (few $\GeV$) VEV $\langle (1,3,0)_{\chi} \rangle$ severely constrained by the measured electroweak precision $\rho$ parameter.
\end{itemize}

\subsection{Scalar potential with $\irrepsub{5}{H}\oplus \irrepsub{10}{S} 
\oplus \irrepsub{24}{S} \oplus \irrepsub{45}{H}$\label{sec:45-model}}

We are only interested in the part of the renormalizable scalar potential that provides the \SM \ invariant contributions to the scalar spectrum proportional to $v_{\rm GUT}$
\begin{align} \label{eq:V}
V & = V_{24}(\chi) + V_{5}(H, \chi) + V_{10}(\phi,\chi) + V_{45}(\Sigma,\chi) 
 + V_{\rm mix}(H, \chi, \Sigma) \; . 
\end{align}
Here
\begin{align}
V_{24} = & -\tfrac{1}{2} \, m_{\chi}^2 \, \chi^{i}_{j} \chi^{j}_{i}
+ \sqrt{\tfrac{10}{3}} \, \mu_{\chi} \, \chi^{i}_{j} \chi^{j}_{k} \chi^{k}_{i}
+ \tfrac{1}{8} \, \lambda_1 \, \chi^{i}_{j} \chi^{j}_{i} \chi^{k}_{l} \chi^{l}_{k}
+ \tfrac{15}{2} \, \lambda_2 \, \chi^{i}_{j} \chi^{j}_{k} \chi^{k}_{l} \chi^{l}_{i}
\:, \label{eq:V_24} \\
V_{5\phantom{0}} = & \ m_{H}^2 \, H_{i}^* H^{i} 
+ \sqrt{30} \, \mu_{H} \, H_{i}^* \chi^{i}_{j} H^{j}
+ \alpha_1 \, H_{i}^* H^{i} \chi^{j}_{k} \chi^{k}_{j}
+ 30 \, \alpha_2 \, H_{i}^* \chi^{i}_{j} \chi^{j}_{k} H^{k}
\:, \label{eq:V_5} \\
V_{10} = & - m_{\phi}^2 \, \phi^*_{ij} \phi^{ji}
+ 2 \sqrt{30} \, \mu_{\phi} \, \phi_{ij}^* \chi^{j}_{k} \phi^{ki} 
+ \beta_1 \, \phi^*_{ij} \phi^{ji} \chi^{k}_{l} \chi^{l}_{k}
+ 60 \, \beta_2 \, \phi_{ij}^{*} \chi^{j}_{k} \chi^{k}_{l} \phi^{li}
+ \nonumber \\ 
& + 30 \, \beta_3 \, \phi_{ij}^{*} \chi^{j}_{k} \phi^{kl} \chi^{i}_{l}
\:,  \label{eq:V_10}\\
V_{45} = & \ m_{\Sigma}^2 \, \Sigma^{ij}_{k} \Sigma_{ij}^{*k}
+ 4 \sqrt{30} \, \mu_{\Sigma} \, \Sigma^{ij}_{k} \Sigma_{ij}^{*l} \chi^{k}_{l}
+ 8 \sqrt{30} \, \mu_{\Sigma}' \, \Sigma^{ij}_{k} \Sigma_{il}^{*k} \chi^{l}_{j} + \nonumber \\
& + \eta_1 \, \Sigma^{ij}_{k} \Sigma_{ij}^{*k} \chi^{l}_{m} \chi^{m}_{l}
+ 120 \, \eta_2 \, \Sigma^{ij}_{k} \Sigma_{ij}^{*l} \chi^{k}_{m} \chi^{m}_{l}
+ 240 \, \eta_3 \, \Sigma^{ij}_{k} \Sigma_{il}^{*k} \chi^{l}_{m} \chi^{m}_{j} + \nonumber \\
& + \tfrac{48}{5} \, \eta_4 \, \Sigma^{ij}_{k} \Sigma_{im}^{*l} \chi^{k}_{j} \chi^{m}_{l}
+ 120 \, \eta_5 \, \Sigma^{ij}_{k} \Sigma_{im}^{*l} \chi^{m}_{j} \chi^{k}_{l}
+ 120 \, \eta_6 \, \Sigma^{ij}_{k} \Sigma_{lm}^{*k} \chi^{l}_{i} \chi^{m}_{j}
\:, \label{eq:V_45} \\
V_{\rm mix} = & \ \tfrac{12\sqrt{5}}{5} \, \tau \, \Sigma^{ij}_{k} \chi^{k}_{i} H_{j}^*
+ 12\sqrt{2} \, \kappa_1 \, \Sigma^{ij}_{k} \chi^{k}_{i} \chi^{l}_{j} H_{l}^*
+ 12\sqrt{2} \, \kappa_2 \, \Sigma^{ij}_{k} \chi^{k}_{l} \chi^{l}_{i} H_{j}^* 
+ \text{h.c.}
\:, \label{eq:V_mix}
\end{align}
and the summation over one upper and one lower repeating index is 
assumed. The potential contains nine real parameters $\{m_{\chi}$, $\mu_{\chi}$, $m_{H}$, $\mu_{H}$, $m_{\phi}$, $\mu_{\phi}$, $m_{\Sigma}$, $\mu_{\Sigma}$, $\mu_{\Sigma}' \}$ and one complex parameter $\{\tau \}$ with positive dimension of mass. There are additional thirteen real and two complex dimensionless 
parameters $\{ \lambda_1$, $\lambda_2$, $\alpha_1$, $\alpha_2$, $\beta_1$, $\beta_2$, $\beta_3$, $\eta_1$, $\eta_2$, $\eta_3$, $\eta_4$, $\eta_5$, $\eta_6 \}$ and $\{ \kappa_1, \kappa_2 \}$, respectively. The signs and various symmetry factors are introduced for convenience. Note that in the unbroken phase the mass terms $\{-m_{\chi}^2$, $m_{H}^2$, $m_{\phi}^2$, $m_{\Sigma}^2 \}$ represent the squared masses of the corresponding $SU(5)$ representations (with the conventional prefactor $\tfrac{1}{2}$ for the real scalar fields and $1$ for complex scalars).
\\

The spectrum presented in Tables~\ref{tab:common-scalar-spectrum} and~\ref{tab:45_H-model-scalar-spectrum} is computed in the minimum of the scalar potential (the vacuum state) obtained for
\begin{align}
\frac{\partial\langle V \rangle}{\partial v_{\rm GUT}} & \equiv 0 \:,
\end{align}
where
\begin{align}
m_{\chi}^2 & = \tfrac{1}{2} \, v_{\rm GUT} \, (2 \, \mu_{\chi} + (\lambda_1 + 14 \, \lambda_2) v_{\rm GUT}) \:,
\end{align}
and $v_{\rm GUT}$ is kept as a free parameter.
The six massless complex scalar states $(3,2,-\tfrac{5}{6})_{\chi}$ are 
absorbed into longitudinal components of twelve heavy gauge bosons.
\begin{table}[th!]
\caption{The scalar spectrum for the simplest \SU{5} embedding of BPR model with only \irrepsub{5}{H}, \irrepsub{10}{S} and \irrepsub{24}{S} multiplets, which corresponds to setting to zero the parameters in $V_{45}$ and $V_{\rm mix}$. Their masses remain unchanged even after adding \irrepsub{45}{H} or \irrepsub{70}{H} to the particle content. Note that the parameters $\mu_{H}$, $\mu_{\phi}$ and $\mu_{\chi}$ should be understood as multiplied by $v_{\rm GUT}$ and $\alpha_1$, $\alpha_2$, $\beta_1$, $\beta_2$, $\beta_3$, $\lambda_1$ and $\lambda_2$ by $v_{\rm GUT}^2$, while each of the masses is a sum of the pertinent contributions. For example,
$ m^{2}(3, 1, -1/3)_H = m_{H}^2 - 2\, \mu_{H}\, v_{\rm GUT}
        + (\alpha_1 + 4\alpha_2) v_{\rm GUT}^2 \;.  $
}
\label{tab:common-scalar-spectrum}
\renewcommand{\arraystretch}{1.2}
\setlength{\tabcolsep}{5pt}
\centering
\vskip 0.2cm
\begin{tabular}{c|*{12}{c|}}
\cline{2-13}
\multicolumn{1}{c|}{} & $m^2_{H}$ & $\mu_{H}$ & $\alpha_1$ & $\alpha_2$ & $m^2_{\phi}$ & $\mu_{\phi}$ & $\beta_1$ & $\beta_2$ & $\beta_3$ & $\mu_{\chi}$ & $\lambda_1$ & $\lambda_2$ 
\\
\hline
\multicolumn{1}{|c|}{{$m^2(1,2,+\tfrac{1}{2})_{H}$}} & $1$ & $3$ & $1$ & $9$ & $$ & $$ & $$ & $$ & $$ & $$ & $$ & $$ \\
\hline
\multicolumn{1}{|c|}{$m^2(3,1,-\tfrac{1}{3})_{H}$} & $1$ & $-2$ & $1$ & $4$ & $$ & $$ & $$ & $$ & $$ & $$ & $$ & $$ \\
\hline
\multicolumn{1}{|c|}{$m^2(3,2,+\tfrac{1}{6})_{\phi}$} & $$ & $$ & $$ & $$ & $1$ & $-1$ & $-1$ & $-13$ & $6$ & $$ & $$ & $$ \\
\hline
\multicolumn{1}{|c|}{$m^2(\xbar{3},1,-\tfrac{2}{3})_{\phi}$} & $$ & $$ & $$ & $$ & $1$ & $4$ & $-1$ & $-8$ & $-4$ & $$ & $$ & $$ \\
\hline
\multicolumn{1}{|c|}{$m^2(1,1,+1)_{\phi}$} & $$ & $$ & $$ & $$ & $1$ & $-6$ & $-1$ & $-18$ & $-9$ & $$ & $$ & $$ \\
\hline
\multicolumn{1}{|c|}{$m^2(1,1,0)_{\chi}$} & $$ & $$ & $$ & $$ & $$ & $$ & $$ & $$ & $$ & $1$ & $1$ & $14$ \\
\hline
\multicolumn{1}{|c|}{$m^2(1,3,0)_{\chi}$} & $$ & $$ & $$ & $$ & $$ & $$ & $$ & $$ & $$ & $5$ & $$ & $20$ \\
\hline
\multicolumn{1}{|c|}{$m^2(8,1,0)_{\chi}$} & $$ & $$ & $$ & $$ & $$ & $$ & $$ & $$ & $$ & $-5$ & $$ & $5$ \\
\hline
\end{tabular}
\end{table}
\begin{table}[th!]
    \caption{Additional contribution to the scalar spectrum in the vacuum after adding \irrepsub{45}{H} to the $\irrepsub{5}{H} \oplus \irrepsub{10}{S} \oplus \irrepsub{24}{S}$ model. The last two rows represent the mixing between \irrepsub{5}{H} and \irrepsub{45}{H}. Again, the parameters $\mu_{\Sigma}$, $\mu_{\Sigma}'$ and $\tau$ should be multiplied by $v_{\rm GUT}$ and $\eta_1$, $\eta_2$, $\eta_3$, $\eta_4$, $\eta_5$, $\eta_6$, $\kappa_1$ and $\kappa_2$ by $v_{\rm GUT}^2$.}
\label{tab:45_H-model-scalar-spectrum}
\centering
\renewcommand{\arraystretch}{1.2}
\setlength{\tabcolsep}{4pt}
\vskip 0.2cm
\begin{tabular}{c|*{12}{c|}}
\cline{2-13}
\multicolumn{1}{c|}{} & $m^2_{\Sigma}$ & $\mu_{\Sigma}$ & $\mu_{\Sigma}'$ & $\eta_1$ & $\eta_2$ & $\eta_3$ & $\eta_4$ & $\eta_5$ & $\eta_6$ & $\tau$ & $\kappa_1$ & $\kappa_2$ 
\\
\hline
\multicolumn{1}{|c|}{{$m^2(1,2,+\tfrac{1}{2})_{\Sigma}$}} & $1$ & $7$ & $19$ & $1$ & $31$ & $67$ & $3$ & $26$ & $21$ & $$ & $$ & $$ \\
\hline
\multicolumn{1}{|c|}{$m^2(3,1,-\tfrac{1}{3})_{\Sigma}$} & $1$ & $2$ & $-6$ & $1$ & $26$ & $42$ & $4$ & $11$ & $-4$ & $$ & $$ & $$ \\
\hline
\multicolumn{1}{|c|}{$m^2(3,3,-\tfrac{1}{3})_{\Sigma}$} & $1$ & $12$ & $4$ & $1$ & $36$ & $52$ & $$ & $6$ & $-24$ & $$ & $$ & $$ \\
\hline
\multicolumn{1}{|c|}{$m^2(\xbar{3},1,+\tfrac{4}{3})_{\Sigma}$} & $1$ & $-8$ & $24$ & $1$ & $16$ & $72$ & $$ & $-24$ & $36$ & $$ & $$ & $$ \\
\hline
\multicolumn{1}{|c|}{$m^2(\xbar{3},2,-\tfrac{7}{6})_{\Sigma}$} & $1$ & $12$ & $-16$ & $1$ & $36$ & $32$ & $$ & $-24$ & $16$ & $$ & $$ & $$ \\
\hline
\multicolumn{1}{|c|}{$m^2(\xbar{6},1,-\tfrac{1}{3})_{\Sigma}$} & $1$ & $-8$ & $-16$ & $1$ & $16$ & $32$ & $$ & $16$ & $16$ & $$ & $$ & $$ \\
\hline
\multicolumn{1}{|c|}{$m^2(8,2,+\tfrac{1}{2})_{\Sigma}$} & $1$ & $-8$ & $4$ & $1$ & $16$ & $52$ & $$ & $-4$ & $-24$ & $$ & $$ & $$ \\
\hline
\multicolumn{1}{|c|}{{$m^2(1,2,+\tfrac{1}{2})_{\rm mix}$}} & $$ & $$ & $$ & $$ & $$ & $$ & $$ & $$ & $$ & $-3$ & \small{$-3\sqrt{3}$} & \small{$-\sqrt{3}$} \\
\hline
\multicolumn{1}{|c|}{$m^2(3,1,-\tfrac{1}{3})_{\rm mix}$} & $$ & $$ & $$ & $$ & $$ & $$ & $$ & $$ & $$ & {\small $2\sqrt{3}$} & $-4$ & $2$ \\
\hline
\end{tabular}
\end{table}

The two $(1,2,+\tfrac{1}{2})$ representations from \irrepsub{5}{H} and \irrepsub{45}{H} mix to form a SM Higgs doublet responsible for electroweak symmetry breaking. To compute their physical masses one needs to diagonalize the matrix
\begin{align}
& \bpm m^2(1,2,+\tfrac{1}{2})_{H} & m^2(1,2,+\tfrac{1}{2})_{\rm mix} \\ (m^2(1,2,+\tfrac{1}{2})_{\rm mix})^* & m^2(1,2,+\tfrac{1}{2})_{\Sigma} \epm .
\end{align}
Similar diagonalization proceeds for the states $(3,1,-\tfrac{1}{3})$ from $H$ and $\Sigma$. One of the masses needs to be around the weak scale to correspond to the SM Higgs. 
It can as well be fine-tuned to zero,
since in our case we have neglected all the $v_{SM}$ contributions to spectrum.
\\

When both of the Higgs doublets develop a non-vanishing VEVs the Georgi-Jarlskog mechanism can be implemented to account for the observed masses of light fermions.
It is also interesting to note that by excluding the mixing terms ($V_{\rm mix}$ with coefficients $\tau$, $\kappa_1$ and $\kappa_2$), as for example in the scenario without \irrepsub{5}{H} where the Higgs doublet belongs entirely to \irrepsub{45}{H}, the masses of fields from $\Sigma$ are not linearly independent, and the following relation among them holds
\begin{align}
& m^2(1,2,+\tfrac{1}{2})_{\Sigma} -\tfrac{3}{4} \, m^2(3,1,-\tfrac{1}{3})_{\Sigma} -\tfrac{9}{8} \, m^2(3,3,-\tfrac{1}{3})_{\Sigma} -\tfrac{3}{4} \, m^2(\xbar{3},1,+\tfrac{4}{3})_{\Sigma} + \nonumber \\
& +\tfrac{3}{4} \, m^2(\xbar{3},2,-\tfrac{7}{6})_{\Sigma} -\tfrac{3}{8} \, m^2(\xbar{6},1,-\tfrac{1}{3})_{\Sigma} +\tfrac{5}{4} \, m^2(8,2,+\tfrac{1}{2})_{\Sigma} = 0 \:.
\end{align}
However, in the most general case the above expressions for scalar masses are all linearly independent.
One can simplify the spectrum even further by imposing an additional $\mathbb{Z}_2$ symmetry
under which in Eq. (\ref{eq:V_24}) $\mu_{\chi} \to 0$, thus imposing a strong correlation between the weak triplet and the color octet masses 
\begin{align}
m^2(1,3,0)_{\chi} & = 20 \, \lambda_2 \, v_{\rm GUT}^2 = 4 \, m^2(8,1,0)_{\chi} \:, \\
m^2(1,1,0)_{\chi} & = 2 \, m_{\chi}^2 \:.
\end{align}
%

\subsection{Scalar potential with $\irrepsub{5}{H}\oplus \irrepsub{10}{S} \oplus \irrepsub{24}{S} \oplus \irrepsub{70}{H}$\label{sec:70-model}}

As long as we are interested only in the $v_{\rm GUT}$-proportional spectrum, the form of the scalar potential remains unaltered upon replacing $\Sigma^{ij}_{k}$ with $\Omega^{ij}_{k}$ in Eqs.~\eqref{eq:V}--\eqref{eq:V_mix}. The corresponding scalar spectrum is shown in Tables~\ref{tab:common-scalar-spectrum} and~\ref{tab:70_H-model-scalar-spectrum}.

\begin{table}[th!]
    \caption{Additional contribution to the scalar spectrum in the vacuum after adding \irrepsub{70}{H} to the $\irrepsub{5}{H} \oplus \irrepsub{10}{S} \oplus \irrepsub{24}{S}$ model. The last two rows represent the mixing between \irrepsub{5}{H} and \irrepsub{70}{H}. Again, $\mu_{\Omega}$, $\mu_{\Omega}'$ and $\widetilde{\tau}$ should be understood as multiplied by $v_{\rm GUT}$ and $\widetilde{\eta}_1$, $\widetilde{\eta}_2$, $\widetilde{\eta}_3$, $\widetilde{\eta}_4$, $\widetilde{\eta}_5$, $\widetilde{\eta}_6$, $\widetilde{\kappa}_1$ and $\widetilde{\kappa}_2$ by $v_{\rm GUT}^2$.}
\label{tab:70_H-model-scalar-spectrum}
\renewcommand{\arraystretch}{1.2}
\setlength{\tabcolsep}{3pt}
\centering
\vskip 0.2cm
\begin{tabular}{c|*{12}{c|}}
\cline{2-13}
\multicolumn{1}{c|}{} & $m^2_{\Omega}$ & $\mu_{\Omega}$ & $\mu_{\Omega}'$ & $\widetilde{\eta}_1$ & $\widetilde{\eta}_2$ & $\widetilde{\eta}_3$ & $\widetilde{\eta}_4$ & $\widetilde{\eta}_5$ & $\widetilde{\eta}_6$ & $\widetilde{\tau}$ & $\widetilde{\kappa}_1$ & $\widetilde{\kappa}_2$ \\
\hline
\multicolumn{1}{|c|}{{$m^2(1,2,+\tfrac{1}{2})_{\Omega}$}} & $1$ & $2$ & $14$ & $1$ & $26$ & $62$ & $6$ & $16$ & $6$ & $$ & $$ & $$ \\
\hline
\multicolumn{1}{|c|}{$m^2(3,1,-\tfrac{1}{3})_{\Omega}$} & $1$ & $\tfrac{16}{3}$ & $-\tfrac{8}{3}$ & $1$ & $\tfrac{88}{3}$ & $\tfrac{136}{3}$ & $\tfrac{16}{3}$ & $\tfrac{28}{3}$ & $-\tfrac{32}{3}$ & $$ & $$ & $$ \\
\hline
\multicolumn{1}{|c|}{$m^2(3,3,-\tfrac{1}{3})_{\Omega}$} & $1$ & $12$ & $4$ & $1$ & $36$ & $52$ & $$ & $6$ & $-24$ & $$ & $$ & $$ \\
\hline
\multicolumn{1}{|c|}{$m^2(\xbar{3},3,+\tfrac{4}{3})_{\Omega}$} & $1$ & $-8$ & $24$ & $1$ & $16$ & $72$ & $$ & $-24$ & $36$ & $$ & $$ & $$ \\
\hline
\multicolumn{1}{|c|}{$m^2(6,2,-\tfrac{7}{6})_{\Omega}$} & $1$ & $12$ & $-16$ & $1$ & $36$ & $32$ & $$ & $-24$ & $16$ & $$ & $$ & $$ \\
\hline
\multicolumn{1}{|c|}{$m^2(15,1,-\tfrac{1}{3})_{\Omega}$} & $1$ & $-8$ & $-16$ & $1$ & $16$ & $32$ & $$ & $16$ & $16$ & $$ & $$ & $$ \\
\hline
\multicolumn{1}{|c|}{$m^2(8,2,+\tfrac{1}{2})_{\Omega}$} & $1$ & $-8$ & $4$ & $1$ & $16$ & $52$ & $$ & $-4$ & $-24$ & $$ & $$ & $$ \\
\hline
\multicolumn{1}{|c|}{$m^2(1,4,+\tfrac{1}{2})_{\Omega}$} & $1$ & $12$ & $24$ & $1$ & $36$ & $72$ & $$ & $36$ & $36$ & $$ & $$ & $$ \\
\hline
\multicolumn{1}{|c|}{{$m^2(1,2,+\tfrac{1}{2})_{\rm mix}$}} & $$ & $$ & $$ & $$ & $$ & $$ & $$ & $$ & $$ & {\small$ -3\sqrt{2}$} & \small{$-3\sqrt{6}$} & \small{$-\sqrt{6}$} \\
\hline
\multicolumn{1}{|c|}{$m^2(3,1,-\tfrac{1}{3})_{\rm mix}$} & $$ & $$ & $$ & $$ & $$ & $$ & $$ & $$ & $$ & $-4$ & $\tfrac{8}{\sqrt{3}}$ & $-\tfrac{4}{\sqrt{3}}$ \\
\hline
\end{tabular}
\end{table}
There are two major differences from the previous case with \irrepsub{45}{H}. As before, disabling the mixing terms in the scalar potential introduces the linear dependence among the masses of \irrepsub{70}{H}
\begin{align}
& m^2(1,2,+\tfrac{1}{2})_{\Omega} - \tfrac{9}{8} \, m^2(3,1,-\tfrac{1}{3})_{\Omega} - \tfrac{3}{8} \, m^2(\xbar{3},3,+\tfrac{4}{3})_{\Omega} + \tfrac{3}{8} \, m^2(6,2,-\tfrac{7}{6})_{\Omega} +
\nonumber \\
& +
\tfrac{1}{4} \, m^2(8,2,+\tfrac{1}{2})_{\Omega} - \tfrac{1}{8} \, m^2(1,4,+\tfrac{1}{2})_{\Omega} = 0 , \\
& m^2(3,3,-\tfrac{1}{3})_{\Omega} + \tfrac{1}{2} \, m^2(\xbar{3},3,+\tfrac{4}{3})_{\Omega} - \tfrac{1}{2} \, m^2(6,2,-\tfrac{7}{6})_{\Omega} + \tfrac{1}{2} \, m^2(15,1,-\tfrac{1}{3})_{\Omega} -
\nonumber \\
& - m^2(8,2,+\tfrac{1}{2})_{\Omega} - \tfrac{1}{2} \, m^2(1,4,+\tfrac{1}{2})_{\Omega} = 0 , \label{eq:linear_dependency_70_eq2}
\end{align}
but now this dependence is preserved even after the states $(1,2,+\tfrac{1}{2})_{\Omega}$ and $(3,1,-\tfrac{1}{3})_{\Omega}$ effectively decouple through the mixing with \irrepsub{5}{H}. As can be seen from Eq.~\eqref{eq:linear_dependency_70_eq2} the rest of the states remain linearly dependent, and since some of them are heavy (e.g. the leptoquarks) a certain fine-tuning is needed to make a particular state light as required by unification.
\\

The second difference comes from the fact that, when considering the full scalar potential for \irrepsub{45}{H} and \irrepsub{70}{H}, they are not of the same form any more due to different symmetry properties of $\Sigma^{ij}_{k}$ and $\Omega^{ij}_{k}$. Namely, since $\phi^{ij}$ is antisymmetric and $\Omega^{ij}_{k}$ is symmetric, all terms contracting $\Omega^{ij}_{k}$ with $\phi_{ij}^*$ vanish. Consequently, the Georgi-Jarlskog mechanism cannot be used in this case and we have to rely on either non-renormalizable Yukawa terms or some other mechanism to explain the pattern of SM fermion masses.

\section{Conclusions}\label{sec:Conclusions}

Although the SM particle set has been completed with the discovery of the Higgs boson in 2012, it is far from being established as unique, isolated  set~\cite{Wells:2017aoy}. 
Our search for possible additional particles proceeds with an aim to both explain the neutrino masses and to achieve the unification of gauge couplings.
With this in mind, we rely on the BSM states employed in the selected Zee-type BPR neutrino model~\cite{Brdar:2013iea}. This set of states allows us to introduce incomplete \SU{5} representations, 
that have a potential to improve gauge coupling crossing. 
Still, this set alone leads to too low unification scale $\MGUT < 10^{15}\,\GeV$, if the Higgs doublet is restricted to belong to \irrepsub{5}{H} irrep. 

In contrast, there are immense possibilities to achieve the successful  unification
if the SM Higgs doublet is embedded into \irrepsub{45}{H}. 
Therefore we specify a plausible set of criteria under which our
search algorithm 
shrinks the number of possibilities to seven successful scenarios listed in Table \ref{tab:H45}. 
In all of them, a light colored scalar $(8,2,1/2)$  provided by \irrepsub{45}{H} plays a decisive role. 
If we choose the SM Higgs belonging to \irrepsub{70}{H} instead
of \irrepsub{45}{H}, our search algorithm selects four scenarios (A1, A3, A4 and A6) from  Table \protect\ref{tab:H45}, and allows for three additional scenarios displayed in Table \ref{tab:H70}. 
Notably, in these new scenarios the
BPR vector-like leptons are assigned to complete irrep \irrepsub{5}{F}, that
do not affect the RGE running. 
Since in these latter scenarios only the scalar \SU{5} irreps are incomplete, 
an eventual verification of them would be in support of a conjecture~\cite{Cox:2016epl} that only scalar irreps may be split.

To conclude, in our procedure of renormalizable \SU{5} embedding, the colorless BPR particles employed in the neutrino mass model get accompanied by the colored partners to enable a successful unification. 
We decide to keep sufficiently heavy those among the colored leptoquark scalars which present a threat to proton stability, and the other colored states may  
play a model-monitoring role both through the LHC
phenomenology~\cite{Giveon:1991zm,Dorsner:2017ufx} and through tests at
Super(and future Hyper)-Kamiokande \cite{Miura:2016krn} experiments.

We also point out that in most of the allowed parameter space the color octet scalar $(8,2,1/2)$ is the most promising BSM state for the LHC searches, and
as such is studied already in \cite{Perez:2016qbo}. 
Additional colored states in the specific gauge unification scenarios 
in Tables~\ref{tab:H45} and \ref{tab:H70} call for a study of characteristic exotic signals at the LHC, which may make some among these specific models falsifiable.

\appendix
\section{Algorithm for optimal unification search}
\label{app:algo}

When studying GUT models with several new BSM states, one
needs a well-defined procedure for identifying the
viable unification scenarios. To this end, it is of some
advantage to ``linearize'' the B-test \eqref{eq:Btest}
\begin{equation}
    \frac{B_{23}}{B_{12}} = 0.718 \equiv b\;,
    \label{eq:defb}
\end{equation}
by first separating the fixed contribution of SM states from 
a contribution of the variable mass BSM states
\begin{equation}
    B_{ij} = B_{ij}^{\rm SM} + B_{ij}^{\rm BSM}\;.
    \label{eq:BijBSM}
\end{equation}
In the next step we write the gauge coupling crossing condition
in the form
\begin{equation}
 B_{23}^{\rm BSM} - b B_{12}^{\rm BSM} =
 b B_{12}^{\rm SM} - B_{23}^{\rm SM} = 1.384 \equiv c \;.
    \label{eq:defc}
\end{equation}
Finally, we separate the $\beta$-function coefficients $\Delta b^{(k)}_i$ from
the threshold weight factors $r_k$  \eqref{eq:rk} of each of $N$ BSM states
and write the crossing test as
\begin{equation}
    \sum_{k=1}^{N} c_k r_k = c \;,
    \label{eq:Ctest}
\end{equation}
where $c_k = (\Delta b_{2}^{(k)} - \Delta b_{3}^{(k)}) 
  - b(\Delta b_{1}^{(k)} - \Delta b_{2}^{(k)})$.
For example, the $E_{L,R}$, $h^+$ and $\Delta$ states
responsible for the neutrino masses in BPR model, if they are
at electroweak scale ($r_k \approx 1$), contribute
to this sum with $\sum  c_k r_k \approx 3$, and significantly overshoot
the required value (\ref{eq:defc}). Thus, we need to add extra states with
total negative contribution $\sum  c_k r_k \approx -1.6$. By choosing
some particular set of states we solve for $r_k$.

In principle, there is an experimental uncertainty of constant $c$
(related to the uncertainty of $b$ in (\ref{eq:defb})), but
we don't need to discuss it because most of the time we will
be able to require the exact gauge crossing, regardless of possible
small variations in the value of $c$.

Along the same lines, expression for the GUT scale \eqref{eq:defMGUT}
can be recast in a condition on $B_{12}^{\rm BSM}$
\begin{equation}
    B_{12}^{\rm BSM} = \frac{184.87}{\ln \MGUT/m_Z} - B_{12}^{\rm SM}
    \approx -1.43 \equiv s \;,
    \label{eq:defs}
\end{equation}
where, to get the numerical value 
we use the low experimental bound on $\MGUT = 5\times 10^{15}
\GeV$. This can again be written in a simple form, linear in
variables $r_k$
\begin{equation}
    \sum_{k=1}^{N} s_k r_k = s \;,
    \label{eq:Stest}
\end{equation}
with $s_k = \Delta b_{1}^{(k)} - \Delta b_{2}^{(k)}$.

Obviously, if we have just one new variable BSM state (N=1) at our
disposal, we can just solve the crossing condition (\ref{eq:Ctest}),
obeying any existing experimental lower bounds
\begin{equation}
    m_{k} \ge m_{k, {\rm min}}  \quad\longrightarrow\quad
    r_{k} \le r_{k, {\rm max}} 
    \label{eq:defrmax}
\end{equation}
and check that GUT scale is high enough.
For two states, $N=2$, $\MGUT$ can also be chosen at will,
and we can either require it equal to experimental lowest
bound, or look for the range of possible $\MGUT$ for which
solution of (\ref{eq:Ctest}) and (\ref{eq:Stest}) exists.
Regardless of this choice, for more than two particles,
$N \ge 3$, the problem becomes under-determined and
we need another criteria. To obtain definite scenarios
we choose to maximize the norm of the vector
\begin{equation}
    \mathbf{r} = (r_1, r_2, \ldots, r_N) \;,
\end{equation}
which means that we choose scenarios with roughly minimal
masses of new particles, or, in other words, we choose
scenarios which have maximal discovery potential.

Following further this principle of maximal discovery potential,
one could also try to minimize at the same time the distance of $\MGUT$ to
the existing experimental lower bound. We have tried this, but
a necessary choice of relative weight of two optimization objectives
brings a complication which we deem unnecessary at this point.
Thus, we performed, for each choice for a set of BSM states, 
a one-dimensional scan with ever increasing fixed $\MGUT$, to find the
range of $\MGUT$ for which the unification scenario works.
The problem can be organized as a standard linear algebra
matrix equation
\begin{equation}
    \mathbf{A} \mathbf{r} \equiv 
 \begin{pmatrix}
     c_1 & c_2  & \dots & c_N \\
     s_1 & s_2  & \dots & s_N 
 \end{pmatrix}
  \begin{pmatrix}r_1 \\ r_2 \\ \hdotsfor{1} \\ r_N \end{pmatrix}
 = \begin{pmatrix} c \\ s \end{pmatrix} \equiv \mathbf{a} \;,
    \label{eq:mateq}
\end{equation}
and if we are temporary not concerned with bounds on $r_k$,
it can be solved using Lagrange multiplier method to
obtain a solution with extremal $\lVert \mathbf{r} \rVert$,
which is $\mathbf{r} = \mathbf{A}^{\tran}(\mathbf{A}
\mathbf{A}^{\tran})^{-1}\mathbf{a}$.
One can also choose to make a variable change
$r_k \to x_k \equiv r_{k, {\rm max}} - r_k$ and minimize
$\lVert \mathbf{x} \rVert$ instead of maximizing 
$\lVert \mathbf{r} \rVert$.
To take the bounds on $r_k$ properly into account, more
sophisticated optimization algorithm is needed and we
used the sequential least squares programming algorithm
SLSQP \cite{Kraft1988}.

\section{Details of SU(5) representations}
\label{app:irreps}

In this Appendix we present the structure and normalizations of used \SU{5} representations.

\paragraph{Adjoint representation}
\scriptsize
\begin{align}
\text{\normalsize$\chi^{i}_{j}$} & = 
\bpm 
-\tfrac{2}{\sqrt{30}} \sigma + \tfrac{1}{\sqrt{2}} O_1 + \tfrac{1}{\sqrt{6}} O_2 & \hspace{-.5cm} O_{R\bar{G}} & \hspace{-.5cm} O_{R\bar{B}} & \hspace{-.5cm} X_R & \hspace{-.5cm} Y_R \\
O_{G\bar{R}} & \hspace{-.5cm} -\tfrac{2}{\sqrt{30}} \sigma - \tfrac{1}{\sqrt{2}} O_1 + \tfrac{1}{\sqrt{6}} O_2 & \hspace{-.5cm} O_{G\bar{B}} & \hspace{-.5cm} X_G & \hspace{-.5cm} Y_G \\
O_{B\bar{R}} & \hspace{-.5cm} O_{B\bar{G}} & \hspace{-.5cm} -\tfrac{2}{\sqrt{30}} \sigma - \tfrac{2}{\sqrt{6}} O_2 & \hspace{-.5cm} X_B & \hspace{-.5cm} Y_B \\
\xbar{X}_{\bar{R}} & \hspace{-.5cm} \xbar{X}_{\bar{G}} & \hspace{-.5cm} \xbar{X}_{\bar{B}} & \hspace{-.5cm} \tfrac{3}{\sqrt{30}} \sigma + \tfrac{1}{\sqrt{2}} \Delta_0 & \hspace{-.5cm} \Delta_+ \\
\xbar{Y}_{\bar{R}} & \hspace{-.5cm} \xbar{Y}_{\bar{G}} & \hspace{-.5cm} \xbar{Y}_{\bar{B}} & \hspace{-.5cm} \Delta_- & \hspace{-.5cm} \tfrac{3}{\sqrt{30}} \sigma - \tfrac{1}{\sqrt{2}} \Delta_0
\epm
\;,
\end{align}
\normalsize 
is the adjoint $24$-dimensional real traceless representation 
\begin{align}
\sum_{i=1}^{5} \chi^{i}_{i} = 0 ,
\end{align}
which  decomposes under the SM group as
\begin{align}
    \irrep{24} = & \overbrace{(1,1,0)}^{\sigma} \oplus \overbrace{(1,3,0)}^{\Delta} \oplus \overbrace{(8,1,0)}^{O} \oplus \overbrace{(3,2,-\tfrac{5}{6})}^{X,Y} \oplus \overbrace{(\xbar{3},2,+\tfrac{5}{6})}^{\bar{X}, \bar{Y}} .
\end{align}
The symbols $\sigma$, $\Delta$, $O$, $X$, $Y$ (and their complex conjugates $\xbar{X}$ and $\xbar{Y}$) denote the SM singlet field, the weak triplet, the color octet and the lower and upper components of $\SU{2}_L$ doublet $(3,2,-\tfrac{5}{6})$, respectively. Note that the singlet $\sigma$, the electrically neutral triplet component $\Delta_0$ and colorless octet components $O_1$ and $O_2$ are real fields so that their mass terms come with a prefactor $\tfrac{1}{2}$. 

\paragraph{Fundamental representation}
$H^{i}$ and $H_{i}^*$ are the $5$-dimensional fundamental and antifundamental complex representations with SM decomposition
\begin{align}
    \irrep{5} = & \overbrace{(1,2,+\tfrac{1}{2})}^{H} \oplus \overbrace{(3,1,-\tfrac{1}{3})}^{S_1} ,
\end{align}
whose fields have the same quantum numbers as
\begin{align}
    \irrepbarsub{5}{F} & = \bpm d^c_{\alpha} \\ \epsilon_{ab} L^b \epm 
= \bpm d^c_{\bar{R}} \\ d^c_{\bar{G}} \\ d^c_{\bar{B}} \\ e \\ -\nu \epm .
\end{align}
and where the weak doublet can play the role of Standard Model Higgs and potentially mix with its counterpart from $\Sigma^{ij}_{k}$ or $\Omega^{ij}_{k}$.

\paragraph{$2$-index antisymmetric representation} $\phi^{ij}$ and $\phi_{ij}^*$ are $10$-dimensional complex antisymmetric representations
\begin{align}
\phi^{ij} & = -\phi^{ji} ,
\end{align}
with the SM decomposition
\begin{align}
    \irrep{10} = & \overbrace{(1,1,+1)}^{h^+} \oplus (\xbar{3},1,-\tfrac{2}{3}) \oplus (3,2,+\tfrac{1}{6}) ,
\end{align}
whose fields have the same quantum numbers as
\begin{align}
    \irrepsub{10}{F} & = \tfrac{1}{\sqrt{2}} \bpm \epsilon^{\alpha\beta\gamma} u^c_{\gamma} & Q^{\alpha b} \\ -(Q^{\beta a})^T & \epsilon^{ab} e^c \epm = \tfrac{1}{\sqrt{2}} 
\bpm 
0 & u^c_{\bar{B}} & -u^c_{\bar{G}} & u_R & d_R \\
-u^c_{\bar{B}} & 0 & u^c_{\bar{R}} & u_G & d_G \\
u^c_{\bar{G}} & -u^c_{\bar{R}} & 0 & u_B & d_B \\
-u_R & -u_G & -u_B & 0 & e^c \\
-d_R & -d_G & -d_B & -e^c & 0
\epm .
\end{align}
The normalization factor $\tfrac{1}{\sqrt{2}}$ is there only for convenience to avoid the double counting of fields in the mass term.

\paragraph{$45$-dimensional representation} $\Sigma^{ij}_{k}$ and $\Sigma_{ij}^{*k}$ are $45$-dimensional complex representations satisfying the antisymmetry and tracelessness conditions
\begin{align}
\Sigma^{ij}_{k} & = -\Sigma^{ji}_{k} \:, \\
\sum_{i=1}^{5} \Sigma^{ij}_{i} & = 0 \:, \qquad  j=1, \ldots , 5 .
\end{align}
Under the SM it is decomposed as
\begin{align}
    \irrep{45} = & \overbrace{(1,2,+\tfrac{1}{2})}^{\Sigma_a} \oplus \overbrace{(3,1,-\tfrac{1}{3})}^{\Sigma_b} \oplus \overbrace{(3,3,-\tfrac{1}{3})}^{\Sigma_c} \oplus \overbrace{(\xbar{3},1,+\tfrac{4}{3})}^{\Sigma_d} \oplus \overbrace{(\xbar{3},2,-\tfrac{7}{6})}^{\Sigma_e} \oplus \nonumber \\ 
& \oplus \underbrace{(\xbar{6},1,-\tfrac{1}{3})}_{\Sigma_f} \oplus \underbrace{(8,2,+\tfrac{1}{2})}_{\Sigma_g} ,
\end{align}
where its non-zero components are
\begin{align}
&
\Sigma^{15}_{4} = -\Sigma^{51}_{4} \to \frac{\Sigma_c^{-R}}{\sqrt{2}} \, , \quad
\Sigma^{25}_{4} = -\Sigma^{52}_{4} \to \frac{\Sigma_c^{-G}}{\sqrt{2}} \, , \quad
\Sigma^{35}_{4} = -\Sigma^{53}_{4} \to \frac{\Sigma_c^{-B}}{\sqrt{2}} \, , \nonumber \\
& 
\Sigma^{14}_{5} = -\Sigma^{41}_{5} \to \frac{\Sigma_c^{+R}}{\sqrt{2}} \, , \quad
\Sigma^{24}_{5} = -\Sigma^{42}_{5} \to \frac{\Sigma_c^{+G}}{\sqrt{2}} \, , \quad
\Sigma^{34}_{5} = -\Sigma^{43}_{5} \to \frac{\Sigma_c^{+B}}{\sqrt{2}} \, ,
\nonumber \\ 
&
\Sigma^{45}_{1} = -\Sigma^{54}_{1} \to \frac{\Sigma_d^{\bar{R}}}{\sqrt{2}} \, , \quad
\Sigma^{45}_{2} = -\Sigma^{54}_{2} \to \frac{\Sigma_d^{\bar{G}}}{\sqrt{2}} \, , \quad
\Sigma^{45}_{3} = -\Sigma^{54}_{3} \to \frac{\Sigma_d^{\bar{B}}}{\sqrt{2}} \, , \nonumber \\
&
\Sigma^{23}_{4} = -\Sigma^{32}_{4} \to \frac{\Sigma_e^{-\bar{R}}}{\sqrt{2}} \, , \quad
\Sigma^{13}_{4} = -\Sigma^{31}_{4} \to \frac{\Sigma_e^{-\bar{G}}}{\sqrt{2}} \, , \quad
\Sigma^{12}_{4} = -\Sigma^{21}_{4} \to \frac{\Sigma_e^{-\bar{B}}}{\sqrt{2}} \, , \nonumber \\
&
\Sigma^{23}_{5} = -\Sigma^{32}_{5} \to \frac{\Sigma_e^{+\bar{R}}}{\sqrt{2}} \, , \quad
\Sigma^{13}_{5} = -\Sigma^{31}_{5} \to \frac{\Sigma_e^{+\bar{G}}}{\sqrt{2}} \, , \quad
\Sigma^{12}_{5} = -\Sigma^{21}_{5} \to \frac{\Sigma_e^{+\bar{B}}}{\sqrt{2}} \, , \nonumber \\
&
\Sigma^{23}_{1} = -\Sigma^{32}_{1} \to \frac{\Sigma_f^{\bar{R}\bar{R}}}{\sqrt{2}} \, , \quad
\Sigma^{13}_{2} = -\Sigma^{31}_{2} \to \frac{\Sigma_f^{\bar{G}\bar{G}}}{\sqrt{2}} \, , \quad
\Sigma^{12}_{3} = -\Sigma^{21}_{3} \to \frac{\Sigma_f^{\bar{B}\bar{B}}}{\sqrt{2}} \, , \nonumber \\
&
\Sigma^{34}_{2} = -\Sigma^{43}_{2} \to \frac{\Sigma_g^{+B\bar{G}}}{\sqrt{2}} \, , \quad
\Sigma^{14}_{3} = -\Sigma^{41}_{3} \to \frac{\Sigma_g^{+R\bar{B}}}{\sqrt{2}} \, , \quad
\Sigma^{24}_{1} = -\Sigma^{42}_{1} \to \frac{\Sigma_g^{+G\bar{R}}}{\sqrt{2}} \, , \nonumber \\
&
\Sigma^{24}_{3} = -\Sigma^{42}_{3} \to \frac{\Sigma_g^{+G\bar{B}}}{\sqrt{2}} \, , \quad
\Sigma^{34}_{1} = -\Sigma^{43}_{1} \to \frac{\Sigma_g^{+B\bar{R}}}{\sqrt{2}} \, , \quad
\Sigma^{14}_{2} = -\Sigma^{41}_{2} \to \frac{\Sigma_g^{+R\bar{G}}}{\sqrt{2}} \, , \nonumber \\
&
\Sigma^{35}_{2} = -\Sigma^{53}_{2} \to \frac{\Sigma_g^{-B\bar{G}}}{\sqrt{2}} \, , \quad
\Sigma^{15}_{3} = -\Sigma^{51}_{3} \to \frac{\Sigma_g^{-R\bar{B}}}{\sqrt{2}} \, , \quad
\Sigma^{25}_{1} = -\Sigma^{52}_{1} \to \frac{\Sigma_g^{-G\bar{R}}}{\sqrt{2}} \, , \nonumber \\
&
\Sigma^{25}_{3} = -\Sigma^{52}_{3} \to \frac{\Sigma_g^{-G\bar{B}}}{\sqrt{2}} \, , \quad
\Sigma^{35}_{1} = -\Sigma^{53}_{1} \to \frac{\Sigma_g^{-B\bar{R}}}{\sqrt{2}} \, , \quad
\Sigma^{15}_{2} = -\Sigma^{51}_{2} \to \frac{\Sigma_g^{-R\bar{G}}}{\sqrt{2}} \, , \nonumber \\
&
\Sigma^{12}_{2} = -\Sigma^{21}_{2} \to \frac{\Sigma_b^R}{2 \sqrt{2}}+\frac{\Sigma_f^R}{2}  \, , \quad
\Sigma^{23}_{3} = -\Sigma^{32}_{3} \to \frac{\Sigma_b^G}{2 \sqrt{2}}-\frac{\Sigma_f^G}{2} \, , \nonumber \\
&
\Sigma^{13}_{1} = -\Sigma^{31}_{1} \to -\frac{\Sigma_b^B}{2\sqrt{2}}+\frac{\Sigma_f^B}{2} \, , \quad
\Sigma^{13}_{3} = -\Sigma^{31}_{3} \to \frac{\Sigma_b^R}{2 \sqrt{2}}-\frac{\Sigma_f^R}{2} \, , \nonumber \\
&
\Sigma^{12}_{1} = -\Sigma^{21}_{1} \to -\frac{\Sigma_b^G}{2\sqrt{2}}-\frac{\Sigma_f^G}{2} \, , \quad
\Sigma^{23}_{2} = -\Sigma^{32}_{2} \to -\frac{\Sigma_b^B}{2 \sqrt{2}}-\frac{\Sigma_f^B}{2} \, , \nonumber \\
&
\Sigma^{14}_{4} = -\Sigma^{41}_{4} \to
-\frac{\Sigma_b^R}{2
\sqrt{2}}+\frac{\Sigma_c^{0R}}{2} \, , \quad
\Sigma^{24}_{4} = -\Sigma^{42}_{4} \to -\frac{\Sigma_b^G}{2
\sqrt{2}}+\frac{\Sigma_c^{0G}}{2} \, , \nonumber \\
&
\Sigma^{34}_{4} = -\Sigma^{43}_{4} \to -\frac{\Sigma_b^B}{2 \sqrt{2}}-\frac{\Sigma_c^{0B}}{2} \, , \quad
\Sigma^{15}_{5} = -\Sigma^{51}_{5} \to -\frac{\Sigma_b^R}{2 \sqrt{2}}-\frac{\Sigma_c^{0R}}{2} \, , \nonumber \\
&
\Sigma^{25}_{5} = -\Sigma^{52}_{5} \to -\frac{\Sigma_b^G}{2 \sqrt{2}}-\frac{\Sigma_c^{0G}}{2} \, , \quad
\Sigma^{35}_{5} = -\Sigma^{53}_{5} \to -\frac{\Sigma_b^B}{2
\sqrt{2}}+\frac{\Sigma_c^{0B}}{2} \, , \nonumber \\
&
\Sigma^{14}_{1} = -\Sigma^{41}_{1} \to \frac{\Sigma_a^+}{2\sqrt{6}}+\frac{\Sigma_g^{+S_1}}{2}+\frac{\Sigma_g^{+S_2}}{2 \sqrt{3}} \, , \quad
\Sigma^{15}_{1} = -\Sigma^{51}_{1} \to -\frac{\Sigma_a^-}{2 \sqrt{6}}+\frac{\Sigma_g^{-S_1}}{2}+\frac{\Sigma_g^{-S_2}}{2 \sqrt{3}} \, , \nonumber \\
&
\Sigma^{24}_{2} = -\Sigma^{42}_{2} \to \frac{\Sigma_a^+}{2 \sqrt{6}}-\frac{\Sigma_g^{+S_1}}{2}+\frac{\Sigma_g^{+S_2}}{2 \sqrt{3}} \, , \quad
\Sigma^{25}_{2} = -\Sigma^{52}_{2} \to
-\frac{\Sigma_a^-}{2 \sqrt{6}}-\frac{\Sigma_g^{-S_1}}{2}+\frac{\Sigma_g^{-S_2}}{2
\sqrt{3}} \, , \nonumber \\
&
\Sigma^{34}_{3} = -\Sigma^{43}_{3} \to \frac{\Sigma_a^+}{2 \sqrt{6}}-\frac{\Sigma_g^{+S_2}}{\sqrt{3}} \, , \quad
\Sigma^{35}_{3} = -\Sigma^{53}_{3} \to -\frac{\Sigma_a^-}{2 \sqrt{6}}-\frac{\Sigma_g^{-S_2}}{\sqrt{3}} \, , \nonumber \\
&
\Sigma^{45}_{5} = -\Sigma^{54}_{5} \to \frac{1}{2}
\sqrt{\frac{3}{2}} \Sigma_a^+ \, , \quad
\Sigma^{45}_{4} = -\Sigma^{54}_{4} \to \frac{1}{2}
\sqrt{\frac{3}{2}} \Sigma_a^- \, ,
\end{align}
with the superscript used to distinguish the individual field components in $\Sigma_{a,\ldots , h}$ by indicating the sign of their $SU(2)_L$ and their $SU(3)_c$ quantum numbers (where $R$, $G$ and $B$ stand for $(\tfrac{1}{2},\tfrac{1}{2\sqrt{3}})$, $(-\tfrac{1}{2},\tfrac{1}{2\sqrt{3}})$ and $(0,-\tfrac{1}{\sqrt{3}})$ pairs
)
under the diagonal generators of the corresponding subgroup.
In this notation it is the field $\Sigma_a^{-}$ (a neutral component of $(1,2,+\tfrac{1}{2})_{\Sigma}$) which develops a non-zero VEV $\langle\Sigma_a^{-}\rangle \equiv v_{45}$.

\paragraph{$70$-dimensional representation} $\Omega^{ij}_{k}$ and $\Omega_{ij}^{*k}$ are $70$-dimensional complex representations satisfying the symmetry and tracelessness conditions
\begin{align}
\Omega^{ij}_{k} & = \Omega^{ji}_{k} , \\
\sum_{i=1}^{5} \Omega^{ij}_{i} & = 0 \quad , \, j=1, \ldots , 5 .
\end{align}
Under SM it gets decomposed as
\begin{align}
    \irrep{70} = & \overbrace{(1,2,+\tfrac{1}{2})}^{\Omega_a} \oplus \overbrace{(3,1,-\tfrac{1}{3})}^{\Omega_b} \oplus \overbrace{(3,3,-\tfrac{1}{3})}^{\Omega_c} \oplus \overbrace{(\xbar{3},3,+\tfrac{4}{3})}^{\Omega_d} \oplus \overbrace{(6,2,-\tfrac{7}{6})}^{\Omega_e} \oplus \nonumber \\ 
& \oplus \underbrace{(15,1,-\tfrac{1}{3})}_{\Omega_f} \oplus \underbrace{(8,2,+\tfrac{1}{2})}_{\Omega_g} \oplus \underbrace{(1,4,+\tfrac{1}{2})}_{\Omega_h} ,
\end{align}
where its non-zero components are
\begin{align}
&
\Omega^{11}_{1} \to \frac{\Omega_b^R}{\sqrt{6}}-\frac{\Omega_f^{R_2}}{\sqrt{2}} \, , \;\;
\Omega^{22}_{1} \to \Omega_f^{GG\bar{R}} \, , \;\;
\Omega^{33}_{1} \to \Omega_f^{BB\bar{R}} \, , \;\;
\Omega^{44}_{1} \to \Omega_d^{+\bar{R}} \, , \;\;
\Omega^{55}_{1} \to \Omega_d^{-\bar{R}} \, , \nonumber \\
&
\Omega^{11}_{2} \to \Omega_f^{RR\bar{G}} \, , \;\;
\Omega^{22}_{2} \to
\frac{\Omega_b^G}{\sqrt{6}}-\frac{\Omega_f^{G_2}}{\sqrt{2}} \, , \;\;
\Omega^{33}_{2} \to \Omega_f^{BB\bar{G}} \, , \;\;
\Omega^{44}_{2} \to \Omega_d^{+\bar{G}} \, , \;\;
\Omega^{55}_{2} \to \Omega_d^{-\bar{G}} \, , \nonumber \\
&
\Omega^{11}_{3} \to \Omega_f^{RR\bar{B}} \, , \;\;
\Omega^{22}_{3} \to \Omega_f^{GG\bar{B}} \, , \;\;
\Omega^{33}_{3} \to
\frac{\Omega_b^B}{\sqrt{6}}-\frac{\Omega_f^{B_2}}{\sqrt{2}} \, , \;\;
\Omega^{44}_{3} \to \Omega_d^{+\bar{B}} \, , \;\;
\Omega^{55}_{3} \to \Omega_d^{-\bar{B}} \, , \nonumber \\
&
\Omega^{11}_{4} \to \Omega_e^{-RR} \, , \;\;
\Omega^{22}_{4} \to \Omega_e^{-GG} \, , \;\;
\Omega^{33}_{4} \to \Omega_e^{-BB} \, , \;\;
\Omega^{44}_{4} \to -\frac{\Omega_a^+}{\sqrt{3}}+\frac{\Omega_h^+}{\sqrt{3}} \, , \;\;
\Omega^{55}_{4} \to \Omega_h^{--} \, , \nonumber \\
&
\Omega^{11}_{5} \to \Omega_e^{+RR} \, , \;\;
\Omega^{22}_{5} \to \Omega_e^{+GG} \, , \;\;
\Omega^{33}_{5} \to \Omega_e^{+BB} \, , \;\;
\Omega^{44}_{5} \to \Omega_h^{++} \, , \;\;
\Omega^{55}_{5} \to \frac{\Omega_a^-}{\sqrt{3}}-\frac{\Omega_h^-}{\sqrt{3}} \, , \nonumber
\end{align}   
\begin{align}
&
\Omega^{15}_{4} = \Omega^{51}_{4} \to \frac{\Omega_c^{-R}}{\sqrt{2}} \, , \quad
\Omega^{25}_{4} = \Omega^{52}_{4} \to \frac{\Omega_c^{-G}}{\sqrt{2}} \, , \quad
\Omega^{35}_{4} = \Omega^{53}_{4} \to \frac{\Omega_c^{-B}}{\sqrt{2}} \, , \nonumber \\
& 
\Omega^{14}_{5} = \Omega^{41}_{5} \to \frac{\Omega_c^{+R}}{\sqrt{2}} \, , \quad
\Omega^{24}_{5} = \Omega^{42}_{5} \to \frac{\Omega_c^{+G}}{\sqrt{2}} \, , \quad
\Omega^{34}_{5} = \Omega^{43}_{5} \to \frac{\Omega_c^{+B}}{\sqrt{2}} \, , \nonumber \\
&
\Omega^{45}_{1} = \Omega^{54}_{1} \to \frac{\Omega_d^{0\bar{R}}}{\sqrt{2}} \, , \quad
\Omega^{45}_{2} = \Omega^{54}_{2} \to
\frac{\Omega_d^{0\bar{G}}}{\sqrt{2}} \, , \quad
\Omega^{45}_{3} = \Omega^{54}_{3} \to \frac{\Omega_d^{0\bar{B}}}{\sqrt{2}} \, , \nonumber \\
&
\Omega^{23}_{4} = \Omega^{32}_{4} \to \frac{\Omega_e^{-\bar{R}}}{\sqrt{2}} \, , \quad
\Omega^{13}_{4} = \Omega^{31}_{4} \to
\frac{\Omega_e^{-\bar{G}}}{\sqrt{2}} \, , \quad
\Omega^{12}_{4} = \Omega^{21}_{4} \to \frac{\Omega_e^{-\bar{B}}}{\sqrt{2}} \, , \nonumber \\
&
\Omega^{23}_{5} = \Omega^{32}_{5} \to
\frac{\Omega_e^{+\bar{R}}}{\sqrt{2}} \, , \quad
\Omega^{13}_{5} = \Omega^{31}_{5} \to \frac{\Omega_e^{+\bar{G}}}{\sqrt{2}} \, , \quad
\Omega^{12}_{5} = \Omega^{21}_{5} \to
\frac{\Omega_e^{+\bar{B}}}{\sqrt{2}} \, , \nonumber \\
&
\Omega^{23}_{1} = \Omega^{32}_{1} \to
\frac{\Omega_f^{\bar{R}\bar{R}}}{\sqrt{2}} \, , \quad
\Omega^{13}_{2} = \Omega^{31}_{2} \to
\frac{\Omega_f^{\bar{G}\bar{G}}}{\sqrt{2}} \, , \quad
\Omega^{12}_{3} = \Omega^{21}_{3} \to
\frac{\Omega_f^{\bar{B}\bar{B}}}{\sqrt{2}} \, , \nonumber \\
&
\Omega^{34}_{2} = \Omega^{43}_{2} \to
\frac{\Omega_g^{+B\bar{G}}}{\sqrt{2}} \, , \quad
\Omega^{14}_{3} = \Omega^{41}_{3} \to
\frac{\Omega_g^{+R\bar{B}}}{\sqrt{2}} \, , \quad
\Omega^{24}_{1} = \Omega^{42}_{1} \to
\frac{\Omega_g^{+G\bar{R}}}{\sqrt{2}} \, , \nonumber \\
&
\Omega^{24}_{3} = \Omega^{42}_{3} \to \frac{\Omega_g^{+G\bar{B}}}{\sqrt{2}} \, , \quad
\Omega^{34}_{1} = \Omega^{43}_{1} \to \frac{\Omega_g^{+B\bar{R}}}{\sqrt{2}} \, , \quad
\Omega^{14}_{2} = \Omega^{41}_{2} \to \frac{\Omega_g^{+R\bar{G}}}{\sqrt{2}} \, , \nonumber \\
&
\Omega^{35}_{2} = \Omega^{53}_{2} \to
\frac{\Omega_g^{-B\bar{G}}}{\sqrt{2}} \, , \quad
\Omega^{15}_{3} = \Omega^{51}_{3} \to
\frac{\Omega_g^{-R\bar{B}}}{\sqrt{2}} \, , \quad
\Omega^{25}_{1} = \Omega^{52}_{1} \to
\frac{\Omega_g^{-G\bar{R}}}{\sqrt{2}} \, , \nonumber \\
&
\Omega^{25}_{3} = \Omega^{52}_{3} \to \frac{\Omega_g^{-G\bar{B}}}{\sqrt{2}} \, , \quad
\Omega^{35}_{1} = \Omega^{53}_{1} \to \frac{\Omega_g^{-B\bar{R}}}{\sqrt{2}} \, , \quad
\Omega^{15}_{2} = \Omega^{51}_{2} \to \frac{\Omega_g^{-R\bar{G}}}{\sqrt{2}} \, , \nonumber \\
&
\Omega^{12}_{2} = \Omega^{21}_{2} \to \frac{\Omega_b^R}{2
\sqrt{6}}+\frac{\Omega_f^{R_1}}{2}+\frac{\Omega_f^{R_2}}{2 \sqrt{2}} \, , \quad
\Omega^{23}_{3} = \Omega^{32}_{3} \to \frac{\Omega_b^G}{2
\sqrt{6}}-\frac{\Omega_f^{G_1}}{2}+\frac{\Omega_f^{G_2}}{2 \sqrt{2}} \, , \nonumber \\
&
\Omega^{13}_{1} = \Omega^{31}_{1} \to
\frac{\Omega_b^B}{2 \sqrt{6}}+\frac{\Omega_f^{B_1}}{2}+\frac{\Omega_f^{B_2}}{2 \sqrt{2}} \, , \quad
\Omega^{13}_{3} = \Omega^{31}_{3} \to \frac{\Omega_b^R}{2
\sqrt{6}}-\frac{\Omega_f^{R_1}}{2}+\frac{\Omega_f^{R_2}}{2 \sqrt{2}} \, , \nonumber \\
&
\Omega^{12}_{1} = \Omega^{21}_{1} \to
\frac{\Omega_b^G}{2 \sqrt{6}}+\frac{\Omega_f^{G_1}}{2}+\frac{\Omega_f^{G_2}}{2 \sqrt{2}} \, , \quad
\Omega^{23}_{2} = \Omega^{32}_{2} \to \frac{\Omega_b^B}{2
\sqrt{6}}-\frac{\Omega_f^{B_1}}{2}+\frac{\Omega_f^{B_2}}{2 \sqrt{2}} \, , \nonumber \\
&
\Omega^{14}_{4} = \Omega^{41}_{4} \to
-\frac{\Omega_b^R}{\sqrt{6}}+\frac{\Omega_c^{0R}}{2} \, , \quad
\Omega^{24}_{4} = \Omega^{42}_{4} \to -\frac{\Omega_b^G}{\sqrt{6}}+\frac{\Omega_c^{0G}}{2} \, , \nonumber \\
&
\Omega^{34}_{4} = \Omega^{43}_{4} \to -\frac{\Omega_b^B}{\sqrt{6}}-\frac{\Omega_c^{0B}}{2} \, , \quad
\Omega^{15}_{5} = \Omega^{51}_{5} \to -\frac{\Omega_b^R}{\sqrt{6}}-\frac{\Omega_c^{0R}}{2} \, , \nonumber \\
&
\Omega^{25}_{5} = \Omega^{52}_{5} \to -\frac{\Omega_b^G}{\sqrt{6}}-\frac{\Omega_c^{0G}}{2} \, , \quad
\Omega^{35}_{5} = \Omega^{53}_{5} \to -\frac{\Omega_b^B}{\sqrt{6}}+\frac{\Omega_c^{0B}}{2} \, , \nonumber \\
&
\Omega^{14}_{1} = \Omega^{41}_{1} \to
\frac{\Omega_a^+}{2 \sqrt{3}}+\frac{\Omega_g^{+S_1}}{2}+\frac{\Omega_g^{+S_2}}{2 \sqrt{3}} \, , \quad
\Omega^{15}_{1} = \Omega^{51}_{1} \to -\frac{\Omega_a^-}{2
\sqrt{3}}+\frac{\Omega_g^{-S_1}}{2}+\frac{\Omega_g^{-S_2}}{2 \sqrt{3}} \, , \nonumber \\
&
\Omega^{24}_{2} = \Omega^{42}_{2} \to \frac{\Omega_a^+}{2
\sqrt{3}}-\frac{\Omega_g^{+S_1}}{2}+\frac{\Omega_g^{+S_2}}{2 \sqrt{3}} \, , \quad
\Omega^{25}_{2} = \Omega^{52}_{2} \to -\frac{\Omega_a^-}{2
\sqrt{3}}-\frac{\Omega_g^{-S_1}}{2}+\frac{\Omega_g^{-S_2}}{2 \sqrt{3}} \, , \nonumber \\
&
\Omega^{34}_{3} = \Omega^{43}_{3} \to \frac{\Omega_a^+}{2
\sqrt{3}}-\frac{\Omega_g^{+S_2}}{\sqrt{3}} \, , \quad
\Omega^{35}_{3} = \Omega^{53}_{3} \to
-\frac{\Omega_a^-}{2 \sqrt{3}}-\frac{\Omega_g^{-S_2}}{\sqrt{3}} \, , \nonumber \\
&
\Omega^{45}_{5} = \Omega^{54}_{5} \to -\frac{\Omega_a^+}{2 \sqrt{3}}-\frac{\Omega_h^+}{\sqrt{3}} \, , \quad
\Omega^{45}_{4} = \Omega^{54}_{4} \to \frac{\Omega_a^-}{2 \sqrt{3}}+\frac{\Omega_h^-}{\sqrt{3}} \, ,
\end{align}
and the same notation for individual field components was used as for $\Sigma^{ij}_{k}$ (with exception of the weak quartet $(1,4,+\tfrac{1}{2})$ fields $\Omega_h^{--}$, $\Omega_h^{-}$, $\Omega_h^{+}$ and $\Omega_h^{++}$ denoted with respect to their increasing $SU(2)_L$ quantum number $-\tfrac{3}{2}$, $-\tfrac{1}{2}$, $+\tfrac{1}{2}$ and $+\tfrac{3}{2}$).
In the chosen notation the neutral components of $(1,2,+\tfrac{1}{2})_{\Omega}$ and $(1,4,+\tfrac{1}{2})_{\Omega}$ can develop a weak-scale VEVs $\langle \Omega_a^{-}\rangle$ and $\langle \Omega_h^{-}\rangle$, respectively.

\subsubsection*{Note}

In the interest of open and reproducible research, computer code
used in production of numerical results for this paper is made available at 
\url{https://github.com/openhep/kmp17}.

\subsubsection*{Acknowledgment}
We would like to thank Borut Bajc, Ilja Dor\v{s}ner, Michal Malinsk\'y and Vasja Susi\v{c} for 
helpful discussions and correspondence at various stages of development of this project.
This work is supported by the Croatian Science Foundation under the project number 8799 
and by the QuantiXLie Center of Excellence KK.01.1.1.01.0004.
IP also acknowledges partial support by the Croatian Science Foundation project
number 4418.



\providecommand{\href}[2]{#2}\begingroup\raggedright\endgroup
\end{document}